\documentclass[twocolumn,showpacs,superscriptaddress,aps,prl]{revtex4-1}
\usepackage{amssymb}
\usepackage{amsmath}
\usepackage{graphicx}
\usepackage{bm}
\usepackage{epsfig}
\usepackage{epstopdf}
\usepackage{amstext}
\usepackage{float}

\begin{document}

\title{Specular Interband Andreev Reflections in Graphene}

\author{D. K. Efetov}
\email{defetov@mit.edu}
\affiliation{Department of Physics, Columbia University, New York, NY 10027, USA}
\author{L. Wang}
\affiliation{Department of Mechanical Engineering, Columbia University, New York, NY
10027, USA}
\author{C. Handschin}
\affiliation{Department of Physics, Columbia University, New York, NY 10027, USA}
\author{K. B. Efetov}
\affiliation{Theoretische Physik III, Ruhr-Universit\"{a}t Bochum, D-44780 Bochum, Germany}
\affiliation{National University of Science and Technology ``MISiS'', Moscow, 119049,
Russia}
\author{J. Shuang}
\affiliation{Department of Chemistry, Princeton University, Princeton, NJ 08544, USA}
\author{R. Cava}
\affiliation{Department of Chemistry, Princeton University, Princeton, NJ 08544, USA}
\author{T. Taniguchi}
\affiliation{National Institute for Materials Science, Namiki 1-1, Ibaraki 305-0044, Japan}
\author{K. Watanabe}
\affiliation{National Institute for Materials Science, Namiki 1-1, Ibaraki 305-0044, Japan}
\author{J. Hone}
\affiliation{Department of Mechanical Engineering, Columbia University, New York, NY
10027, USA}
\author{C. R. Dean}
\affiliation{Department of Physics, Columbia University, New York, NY 10027, USA}
\author{P. Kim}
\email{pk2015@columbia.edu}
\affiliation{Department of Physics, Columbia University, New York, NY 10027, USA}
\date{\today}

\begin{abstract}
Electrons incident from a normal metal onto a superconductor are reflected back as holes -- a process called Andreev reflection~\cite{AR1,BdG,SC1}. In a normal metal where the Fermi energy is much larger than a typical superconducting gap, the reflected hole retraces the path taken by the incident electron. In graphene with ultra low disorder, however, the Fermi energy can be tuned to be smaller than the superconducting gap. In this unusual limit, the holes are expected to be reflected specularly at the superconductor-graphene interface due to the onset of interband Andreev processes, where the effective mass of the reflected holes change sign~\cite{SAR1,SAR2}. Here we present measurements of gate modulated Andreev reflections across the low disorder van der Waals interface formed between graphene and the superconducting NbSe$_{2}$. We find that the conductance across the graphene-superconductor interface exhibits a characteristic suppression when the Fermi energy is tuned to values smaller than the superconducting gap, a hallmark for the transition between intraband retro- and interband specular- Andreev reflections.
\end{abstract}

\pacs{78.30.cd,75.47.Lx,75.70.Tj,73.50.Jt}
\maketitle

Andreev reflection (AR) is a process of transferring charges from a normal metal (N) into a superconductor (SC)~\cite{AR1,BdG,SC1}. When an NS interface is biased with an energy $eV_{ns}$ above the Fermi energy $\varepsilon_{F}$, an electron can only be injected into the SC gap $\Delta$, if a hole is reflected back with an energy of $eV_{ns}$ below $\varepsilon _{F}$, creating a Cooper pair at the Fermi level. For a straight N/SC interface, the momentum conservation along the boundary must be conserved. Thus the incident angle of an incoming electron $\theta_{inc}$ and the reflected angle of a hole, $\theta_{ref}$, have a simple relation $p_e \sin \theta_{inc}=p_h \sin \theta_{ref}$, where $p_e$ and $p_h$ are the total momentum of the electron and hole, respectively. In the limit, $\varepsilon_{F}\gg \Delta$, which holds for a typical NS junction, the reflected hole remains in the conduction band of the  normal metal,  and therefore necessarily carries opposite sign of the mass compared with the electron.  In order to conserve momentum the hole reflects back along a path of the incident electron, exhibiting nearly perfect retro- AR, with $\theta_{ref}\approx -\theta_{inc}$~\cite{AR2,AR3,BKT1}.

If however the Fermi energy can be tuned such that $\varepsilon_{F} \leq \Delta$, a completely different kind of AR is expected.  In this case the energy difference, $2eV_{ns}$, provided by the AR process can result in the reflected hole appearing in the valence band rather than the conduction band.  The reflected hole now has the same mass sign as the incident electron, and therefore according to momentum conservation, travels in the same direction along the interface ($\theta_{ref} > 0$ in Fig. 1). In the ultimate limit of $\varepsilon _{F}$=$0$, the angle of reflection equals to the angle of incidence $\theta _{ref}$=$\theta_{inc}$ - a process called specular reflection~\cite{SAR1,SAR2}.  The condition for specular AR is satisfied whenever $\Delta > eV_{ns} > \varepsilon _{F}$ (Fig. 1(a) right panel), and is therefore predicted to be observable when the normal metal in the NS junction consists of a zero-gap semiconductor (ZGS), and the Fermi energy is tuned close to the charge neutrality point (CNP) where the conduction band and valence band meet.

\begin{figure*}[tbp]
\centering
\includegraphics[width=1\textwidth]{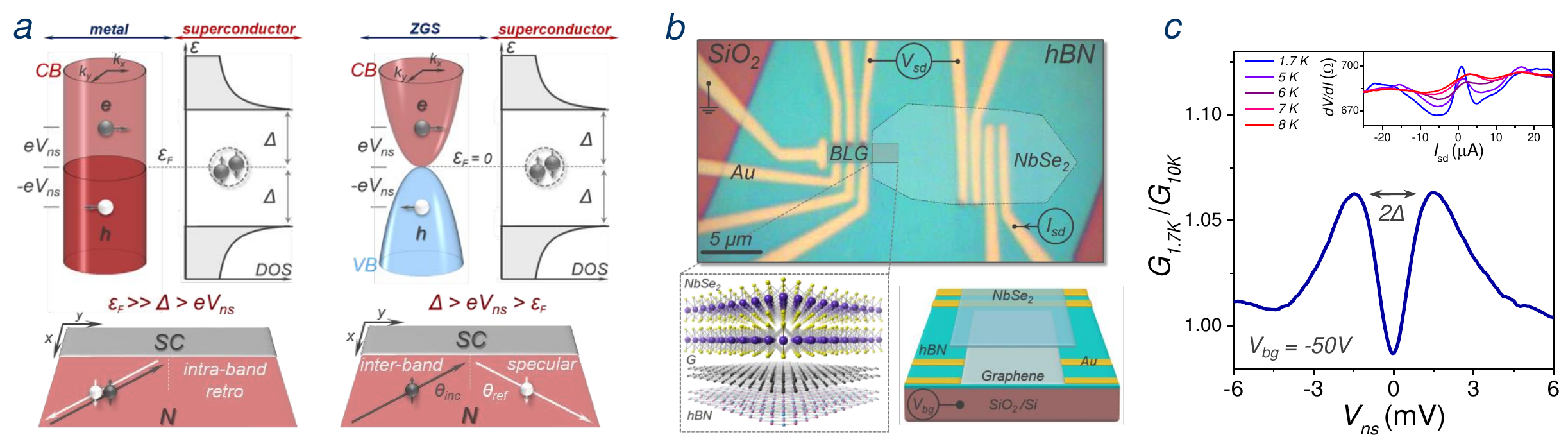}
\caption[Andreev reflections in hBN/BLG/NbSe$_{2}$ hetero-structures]
{(a) Schematics of the AR processes at a normal metal/SC (left) and a zero-gap semiconductor/SC interface (right). An electron from the CB with a total energy of $\protect\varepsilon_{F}+eV_{ns}$ is reflected by a hole with an energy $\protect\varepsilon_{F}-eV_{ns}$, forming a Cooper pair at $\protect\varepsilon_{F}$ in the SC. For large $\protect\varepsilon_{F}$ the hole remains in the CB, resulting in an intraband retro-reflection process. For small enough $\protect\varepsilon_{F}$ the hole undergoes an interband transition into the VB resulting in specular-reflections.
(b) Top : color enhanced optical image of the hBN/BLG device before the transfer of the NbSe$_{2}$ flake (marking its final position). The NbSe$_{2}$ and the hBN/BLG stack form an electrically coupled overlap region. Bottom left : schematic diagram of the vertical cross-section of the vdW stack. Bottom right : sketch of the final device on top of a $300$~nm SiO$_{2}$/Si back gate.
(c) Inset : temperature dependence of differential resistance $dV/dI$ vs. bias current $I_{sd}$ across the BLG/NbSe$_{2}$ interface at $V_{bg}$=$-50$~V. Strong variation of $dV/dI$ appear below NbSe$_{2}$ $T_{c} \sim 7$~K. Main : normalized conductance $G_{1.7K}/G_{10K}$ vs. $V_{ns}$. The
characteristic double-peak line-shape arises from ARs in the limit $\varepsilon_{F} \gg \Delta$.}
\end{figure*}

Graphene provides an ideal platform to exhibit both intra and interband AR. Its two-dimensional nature and ZGS properties enable to induce a crossover from inter- and intraband AR by tuning $\varepsilon _{F}$ with the electric field effect~\cite{graphene,BLG1,BLG2}. However, accessing the regime $\varepsilon _{F}\sim \Delta$, a necessary condition to realize the interband AR, has been technically challenging. In typical graphene samples on a SiO$_2$ substrate, strong potential fluctuations up to $\delta\varepsilon _{F}\sim 50$~meV have been typically observed due to the presence of charged impurities~\cite{puddles1}. This value is much larger than $\Delta$ in typical SC. The recent progress in producing suspended~\cite{susp} and hBN supported graphene samples~\cite{BN1} has now allowed to dramatically reduce these fluctuations down to $\delta \varepsilon_{F}\sim 5$~meV~\cite{puddles2}. Despite experimental progress has been steadily made in contacting graphene with various superconducting metals~\cite{JEG1,JEG2,JEG3,JEG4,JEG5} including recent work of the edge contact on graphene ~\cite{Delft,Geim}, fabricating transparent SC contacts on graphene channels with extremely low inhomogeneity has yet to be realized.

In this letter we employ a novel non-invasive approach to fabricate N/SC interfaces with an unprecedented energy resolution close to the CNP. For this purpose we electronically couple (see supplementary information (SI) for detail method) a high mobility hBN/bilayer graphene (BLG) device and a $20$-$100$~nm thin NbSe$_{2}$ flake. NbSe$_{2}$ is a van der Waals (vdW) SC with a critical temperature T$_{c}\sim 7$~K and a large $\Delta\sim 1.2$~meV~\cite{NbSe1,NbSe2,NbSe3} (Fig. 1 (b)). We use the dry-vdW transfer technique and a current annealing method~\cite{BN1,vdW1} to achieve ultra-clean, atomically sharp, and highly transparent Ohmic interfaces between graphene and NbSe$_{2}$. In this experiment, we chose BLG rather than single layer graphene to minimize $\delta\varepsilon_{F}$ near the CNP. Owing to the larger density of states of BLG near the CNP, smaller $\delta\varepsilon_{F}$ can be obtained in BLG given the similar level of substrate induced inhomogeneity. The devices were fabricated on heavily degenerated Si substrates topped with $300$~nm SiO$_{2}$, where a back gate voltage $V_{bg}$ is applied to tune $\varepsilon _{F}$ of the BLG channel.

We characterize the electronic transport properties across the vdW N/SC junction by measuring differential resistance $dV/dI$ as a function of the channel current $I_{sd}$ (Fig. 1 (b)). Inset of Fig. 1 (c) shows typical traces for a high back-gate voltage $V_{bg}$=$-50$~V that corresponds to a representative condition $\varepsilon _{F}\gg \Delta$. We observe that the $dV/dI(I_{sd})$ traces become increasingly non-linear for temperatures $T < T_{c}$. In order to resolve these features better, we divide traces taken below and above $T_{c}$ (here at $T=1.7$~K and $T=10$~K) and obtain the normalized differential conductance $G_{1.7K}/G_{10K}$=$(dV/dI_{10K})/(dV/dI_{1.7K})$. Fig. 1 (c) shows $G_{1.7K}/G_{10K}$ as a function of the voltage drop across the N/SC junction $V_{ns}$, estimated by considering the BLG channel resistance (SI). The most salient feature of the resulting $G_{1.7K}/G_{10K}(V_{ns})$ curve is a conductance dip around zero bias with two pronounced conductance peaks at $|V_{ns}|\sim \pm$1.2~mV. The position of these peaks is consistent with the value of $\Delta _{NbSe_{2}}$. Later, we will show that this line shape is characteristic for typical N/SC junctions with $\varepsilon _{F}\gg \Delta $ and can be explained in terms of intraband retro-ARs~\cite{BKT1,AR2,AR3}.

The ability to tune $\varepsilon_{F}$ enables us to investigate the characteristic AR signal at the continuous transition from large to small $\varepsilon_{F}$. Fig. 2 (a) left panel shows the characteristic longitudinal resistance $R_{xx}$ of the BLG channel vs. $V_{bg}$. An upper bound of $\delta \varepsilon_{F} < 1$~meV was estimated from the full-width-half-maximum (FWHM) peak of $R_{xx}(V_{bg})$~\cite{BN1,susp,puddles2,xxx}, demonstrating that the condition $\varepsilon_{F} < \Delta$ can be realized in this sample. Fig. 2(a) right panel displays simultaneous measurements of $G_{1.7K}/G_{10K}$ as a function of both the bias current $I_{sd}$ and $V_{bg}$. For large $\varepsilon_{F}$, $|V_{bg}| > 2$~V, $G_{1.7K}/G_{10K}(I_{sd})$ exhibits a similar characteristic behaviour described in Fig. 1(c), showing the characteristic conductance dip around the zero bias condition $I_{sd}=0$. However near the CNP, for $|V_{bg}| < 2$~V, the $G_{1.7K}/G_{10K}(I_{sd})$ traces exhibit drastic variations : the conductance dip at zero bias turns into a peak and then is suppressed back again as $\varepsilon_{F}$ goes through the CNP.

\begin{figure}[tbp]
\centering
\includegraphics[width=0.5\textwidth]{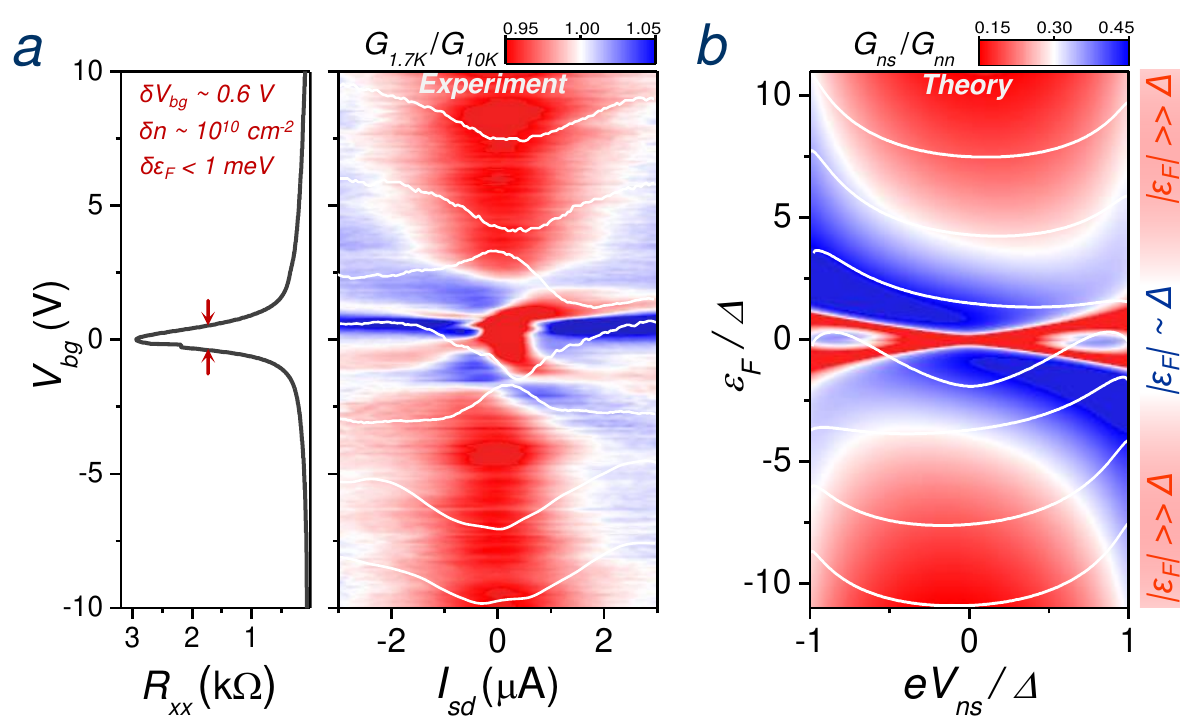}
\caption[Gate tunable Andreev reflections]
{(a) Left : $R_{xx}$ vs. $V_{bg}$ showing the characteristic channel resistance peak at the CNP with a full-width of half-maximum of $\protect\delta V_{bg} \sim 0.6$~V, suggesting an upper bound of $\delta \varepsilon_{F}<$1~meV. Right : color map of normalized differential conductance $G_{1.7K}/G_{10K}$ as a function of bias current $I_{sd}$ and the gate voltage $V_{bg}$ for the inner gap region $|eV_{ns}|<\Delta$ with line-traces of $G_{1.7K}/G_{10K}(I_{sd})$ for various fixed $V_{bg}$.
(b) Color map of the theoretically obtained normalized differential conductance $G_{ns}/G_{nn}$ vs. $\varepsilon_{F}/\Delta$ and $V_{ns}/\Delta$ with line-traces of $G_{ns}/G_{nn}(eV_{ns}/\Delta)$ for various fixed $\varepsilon_{F}/\Delta$, showing overall agreement with the experimental data. The gate tunability of BLG allows to continuously probe the conductance changes from the large $|\protect\varepsilon_{F}| \gg \Delta$ to the small $|\protect\varepsilon_{F}|\sim\Delta$ limit. The zero bias dip line-shape at large $\protect\varepsilon_{F}$ is continuously transformed into a diagonal cross-like line-shape close to the CNP.}
\end{figure}

To explain these experimental findings at the crossover from the large to the small $\varepsilon_{F}$ limit, we develop a theoretical model based on the Bogoliubov-deGennes equations and the Blonder-Tinkham-Klapwijk (BTK) formalism for the conductance across the BLG/SC interface $G_{ns}$ at $T$=$0$~K. We also compute the normal conductance $G_{nn}$ for $T > T_{c}$ ($\sim 10$~K) (see SI for details). Fig. 2 (b) shows the resulting normalized conductance $G_{ns}/G_{nn}$ as a function of $\varepsilon _{F}/\Delta$ and $eV_{ns}/\Delta$ for comparable energy ranges as the experimental data. The theoretical model demonstrates qualitatively good agreement with the experimentally obtained normalized conductance in Fig. 2(a). Both the broad variation of the normalized conductance in the regime of $|\varepsilon _{F}|\gg \Delta $ and the rapid variation in the regime of $|\varepsilon_{F}|\sim \Delta$ are matched well with our experimental observation.

Further quantitative comparison between the experiment and theoretical model for $|\varepsilon _{F}| \sim \Delta$ can be performed by re-plotting the experimental $G_{1.7K}/G_{10K}$ map as a function of $\varepsilon _{F}$ and $V_{ns}$ (see SI for the conversion scheme) (Fig. 3 (a)). In both graphs, one can identify four regions of enhanced conductance (colored blue): two of them for $|\varepsilon_{F}| > |eV_{ns}|$ and two for $|\varepsilon _{F}| < |eV_{ns}|$. These regions are separated from each other by connected regions of reduced conductance (colored red) that are approximately following the dependence $|\varepsilon _{F}|\sim |eV_{ns}|$, forming diagonal lines that are roughly symmetrically arranged with respect to $\varepsilon _{F}$ = $eV_{ns}$ = $0$ in the conductance maps. Several representative line cuts, showing $G_{1.7K}/G_{10K}(\varepsilon _{F})$, clearly exhibit similar features for both experimental and theoretical traces (Fig. 3 (b)).

\begin{figure*}[tbp]
\centering
\includegraphics[width=0.7\textwidth]{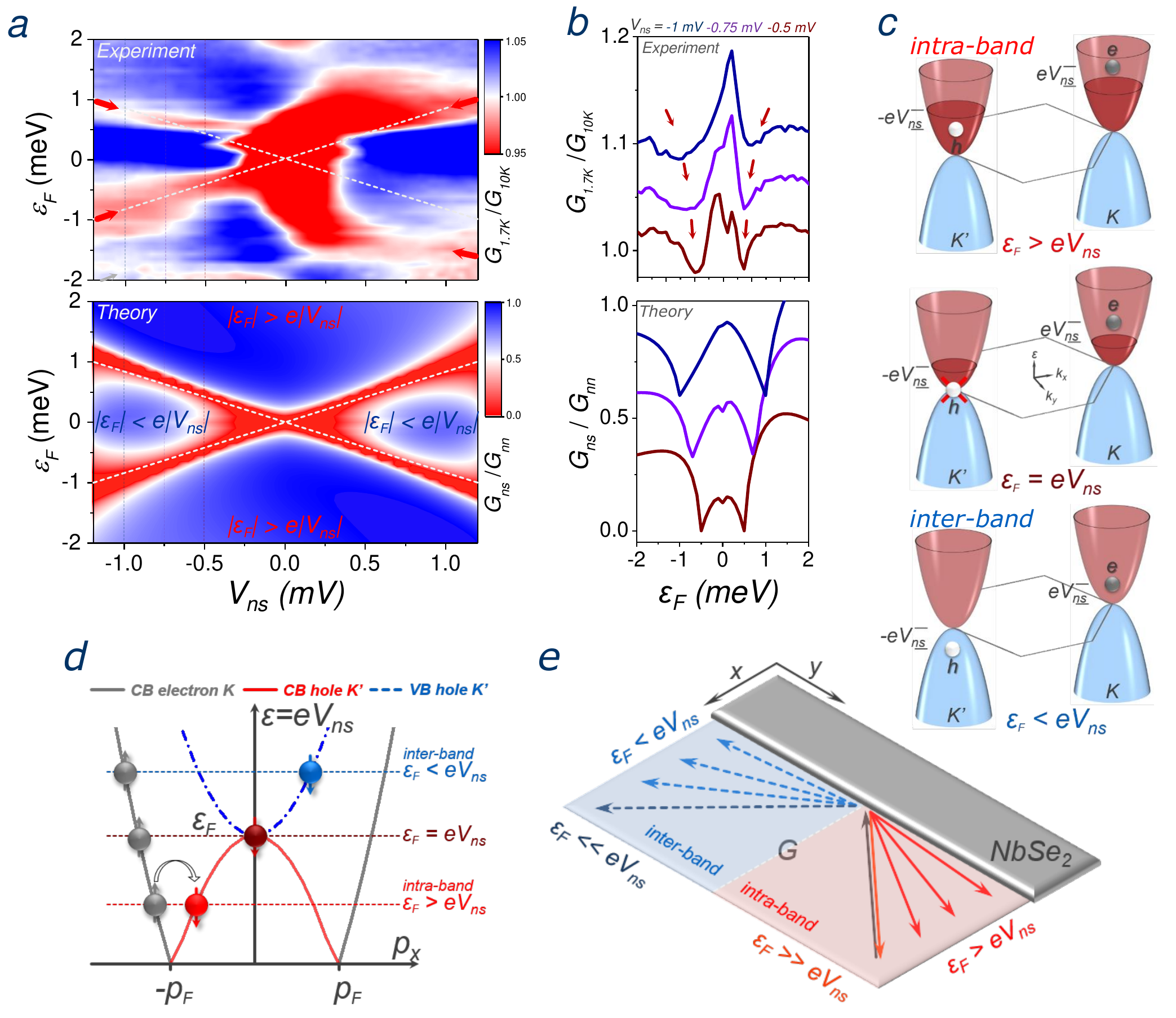}
\caption[Specular interband Andreev reflections close to the CNP]
{(a) Experimental $G_{1.7K}/G_{10K}$ and theoretical $G_{ns}/G_{nn}$ color maps as a function of $V_{ns}$ and $\protect\varepsilon_{F}$ in the limit $|\protect\varepsilon_{F}| \sim \Delta$. A continuous region of lower conductance (red) that is defined for $|\varepsilon_{F}| \sim |eV_{ns}|$ (white dashed lines) subdivides the map into four disconnected regions of comparatively high conductances (blue). In the regions where $\protect|\varepsilon_{F}| > |eV_{ns}|$ the ARs are of the intraband retro type and in the regions where $\protect|\varepsilon_{F}| < |eV_{ns}|$ the ARs are of the interband specular type. (b) Experimental $G_{1.7K}/G_{10K}(\protect\varepsilon_{F})$ and theoretical $G_{ns}/G_{nn}(\protect\varepsilon_{F})$ line-traces demonstrate
the evolution of the conductance dips (red arrows) with varying $V_{ns}$. (c) Schematics of the AR process for BLG at the cross-over from intraband to interband ARs. With decreasing $\protect\varepsilon_{F}$ at a fixed $V_{ns}$ the AR hole moves from the CB to the VB. The cross-over point where the hole is reflected onto the CNP is defined by $\protect\varepsilon_{F}$=$eV_{ns}$. (d) Excitation spectrum $\protect\varepsilon(p_{x})$=$eV_{ns}$ for a fixed $\protect\varepsilon_{F} < \Delta$. With an increasing excitation voltage $V_{ns}$, the momentum $p_{x}$ of the reflected hole continuously increases from negative to positive values, passing through zero when $\protect\varepsilon_{F}$=$eV_{ns}$. (e) Schematics of the reflection angles of AR holes in the various energy limits. Starting from perfect intraband retro-reflections in the high $\protect\varepsilon_{F}$ limit, $\protect\theta_{ref}$ continuously increases toward $\pi/2$ as $\varepsilon_{F}$ is lowered. At the cross-over point separating intraband and interband ARs, $\protect\varepsilon_{F}$=$eV_{ns}$, $\protect\theta_{ref}$ exhibits a jump to $-\pi/2$, which eventually results in perfect interband specular reflections ($\theta_{ref}$=$\theta_{inc}$) when $\protect\varepsilon_{F}$=$0$.}
\end{figure*}

The various features observed in the conductance map can be explained by analyzing the microscopic processes for different $\varepsilon_{F}$ (Fig. 3 (c)). ARs for SLG and BLG involve intervalley processes due to the time reversal symmetry of the backwards motion of the reflected hole~\cite{JEG1,JEG2,SAR1,SAR3}. Therefore, for $\varepsilon _{F} > eV_{ns}$, an electron in the $K$-valley of the CB with an energy of $\varepsilon_{F} + eV_{ns}$ is reflected as a hole in the $K^{\prime }$-valley of the CB with an energy $\varepsilon_{F} - eV_{ns}>0$. This intraband AR process gives rise to a relatively high conductance, analogous to ARs in normal metals in the limits $\varepsilon _{F}\gg \Delta $. When $\varepsilon_{F}$ decreases, however, the phase space for the reflected hole is decreased, resulting in decreasing conductance. This effect culminates in a minimum in conductance at the condition $\varepsilon_{F} = eV_{ns}$, where the hole is at the CNP and intraband AR cease to exist. For $\varepsilon_{F} < eV_{ns}$ the hole undergoes an interband transition into the VB and the conductance again increases.

This non-monotonic conductance change as a function of $\varepsilon_{F}$ can be quantitatively explained by the existence of a critical angle $\theta_{c} \equiv \sin^{-1}\sqrt{|\varepsilon _{F}-eV_{ns}|/|\varepsilon_{F}+eV_{ns}|}$ (SI) that allows AR process only to happen for electrons that are incident at $\theta_{inc}\leq\theta_c$. In the limit of the intraband AR ($\varepsilon _{F}\gg eV_{ns}$) and the interband AR ($\varepsilon _{F}\ll eV_{ns}$), $\theta_{c}\approx \pi/2$, and thus most electrons undergo AR processes resulting in a high conductance. However, for $\varepsilon _{F} \sim eV_{ns}$, the critical angle approaches zero.  If we assume a distribution of angles for the incident electrons, those outside of the critical angle will be reflected normally and only those arriving at near normal incidence will undergo  AR processes, resulting in an overall suppressed conductance across the junction. For $\varepsilon_{F}$=$eV_{ns}$, $\theta_{c}$=$0$, and no electrons can enter the SC. The resulting conductance minima hence mark the exact crossover points between intraband and interband ARs.

We can now connect these processes with the corresponding changes in $\protect\theta_{ref}$ (Fig. 3 (d) and (e)). The dispersion relation $E(p_{x},p_{y})$ of BLG is shown in Fig. 3(d), where $p_{x}$ and $p_{y}$ are momenta perpendicular and parallel to the N/SC interface, respectively. Considering that the parallel momentum $p_{y}$ is conserved in the process~\cite{SAR1,SAR2}, the excitation spectrum for ARs $\varepsilon\equiv eV_{ns}=|E-\varepsilon_{F}|$ can be expressed by a function of only $p_{x}$. Depending on the relative size of energy of the incident CB electron (assuming that it is in the $K$-valley) compared to $\varepsilon_{F}$, we can then relate the reflection process to three different scenarios: (i) Andreev reflected CB holes (normal AR), (ii) all CB electrons are normally reflected (no AR), and (iii) Andreev reflected VB holes (specular AR), all in the $K'$-valley. For small $eV_{ns}\ll \varepsilon_{F}$, $p_{x}$ of the incident electron and the intraband reflected CB hole is almost unchanged. Since a CB hole has a negative mass, its negative $p_{x}$ and the conserved $p_{y}$ provide an overall direction of motion that retraces that of the incident electron (i.e., (i) above). As $V_{ns}$ grows, $p_{x}$ of the reflected hole increases, translating into a larger $\protect\theta_{ref}$, that ultimately becomes $\pi/2$ when $\varepsilon_{F}$ = $eV_{ns}$ and $p_{x}=0$. As $V_{ns}$ grows further, so that $eV_{ns} >  \varepsilon_{F}$, the incident electron is reflected as an interband VB hole, which has a positive mass and a positive $p_{x}$ that is opposite to the sign of $p_{x}$ of the incident electron. While the simultaneous sign changes of the mass and of $p_{x}$ have no effect on the motion in the $x$-direction, the sign change of the mass abruptly reverses the motion in the $y$-direction by $180^{\circ}$ due to the conservation of $p_{y}$, as described in (iii) above. For the ultimate limits $\varepsilon_{F} \gg \Delta$ and  $\varepsilon_{F}$=$0$ one obtains perfect retro-reflections and specular-reflections respectively with varying angles for intermediate energy scales and a discontinuous jump at $eV_{ns}=\varepsilon_{F}$. Summarizing this discussion, the conductance maps in Fig. 3 (a) can now be used as phase diagrams separating regions of intraband retro-AR, where $|\varepsilon_{F}|>|eV_{ns}|$, and interband specular-AR, where $|\varepsilon_{F}| < |eV_{ns}|$.

While overall the experimental findings are in qualitative agreement with the theoretical estimates, there are certain quantitative mismatches. Notably, the experimental conductance map has an overall higher conductance with broadened features. In addition, we observe a pronounced region of reduced conductance around $V_{ns}\approx 0$ for a wide range of $\varepsilon_{F}$ that is not present in the theoretical map. All these deviations can be attributed to imperfections due to a realistically broadened N/SC interface, inelastic scattering at finite temperatures and the presence of small potential fluctuations, whose quantitative descriptions go beyond the scope of our simple theoretical model.

In conclusion we have developed unprecedentedly clean BLG/NbSe$_{2}$ based N/SC junctions that allowed to study ARs at low $\varepsilon_{F}$. Our observation of gate tunable transitions between retro intraband and specular interband ARs opens a new route for future experiments that could employ the gate control of $\protect\theta_{ref}$, which can be continently and independently altered with $V_{bg}$ and $V_{ns}$. Most importantly our finding help to draw a general picture of the exact physical processes underlying ARs.

\begin{acknowledgments}
The authors gratefully acknowledge fruitful discussions with C. Beenakker, I. Aleiner, L. Levitov, A. McDonald, A. Volkov, P. Jarrillo-Herrero, A. Akhmerov, C. Schoenenberger, M. Devoret and J. Pillet. DKE acknowledges the support of the FAME Center, sponsored by SRC MARCO and DARPA. PK acknowledge support by DOE (DE-FG02-05ER46215) and CH is supported by AFOSR (FA2386-13-1-4122). KBE gratefully acknowledges the financial support of the Ministry of Education and Science of the Russian Federation in the framework of Increase Competitiveness Program of NUST \textquotedblleft MISiS\textquotedblright\ (Nr. K2-2014-015) as well as of Transregio 12 and Priority Program 1459 \textquotedblleft Graphene\textquotedblright\ of \textit{Deutsche Forschungsgemeinschaft}. K.W. and T.T. acknowledge support from the Elemental Strategy Initiative conducted by the MEXT, Japan. T.T. acknowledges support a Grant-in-Aid for Scientific Research on Grant 262480621 and on Innovative Areas “Nano Informatics” (Grant 25106006) from JSPS.

\end{acknowledgments}

\bibliographystyle{apsrev}
\bibliography{SAR}

\begin{thebibliography}{34}
\expandafter\ifx\csname natexlab\endcsname\relax\def\natexlab#1{#1}\fi
\expandafter\ifx\csname bibnamefont\endcsname\relax
  \def\bibnamefont#1{#1}\fi
\expandafter\ifx\csname bibfnamefont\endcsname\relax
  \def\bibfnamefont#1{#1}\fi
\expandafter\ifx\csname citenamefont\endcsname\relax
  \def\citenamefont#1{#1}\fi
\expandafter\ifx\csname url\endcsname\relax
  \def\url#1{\texttt{#1}}\fi
\expandafter\ifx\csname urlprefix\endcsname\relax\def\urlprefix{URL }\fi
\providecommand{\bibinfo}[2]{#2}
\providecommand{\eprint}[2][]{\url{#2}}

\bibitem[{\citenamefont{Andreev}(1964)}]{AR1}
\bibinfo{author}{\bibfnamefont{A.~F.} \bibnamefont{Andreev}},
  \bibinfo{journal}{Soviet Physics JETP} \textbf{\bibinfo{volume}{19}},
  \bibinfo{pages}{1228} (\bibinfo{year}{1964}).

\bibitem[{\citenamefont{DeGennes}(1966)}]{BdG}
\bibinfo{author}{\bibfnamefont{P.~G.} \bibnamefont{DeGennes}},
  \emph{\bibinfo{title}{Superconductivity of metals and alloys}}
  (\bibinfo{publisher}{Benjamin}, \bibinfo{address}{New York},
  \bibinfo{year}{1966}).

\bibitem[{\citenamefont{Tinkham}(1996)}]{SC1}
\bibinfo{author}{\bibfnamefont{M.}~\bibnamefont{Tinkham}},
  \emph{\bibinfo{title}{Introduction to superconductivity}}, International
  series in pure and applied physics (\bibinfo{publisher}{McGraw Hill},
  \bibinfo{address}{New York}, \bibinfo{year}{1996}), \bibinfo{edition}{2nd}
  ed.

\bibitem[{\citenamefont{Beenakker}(2006)}]{SAR1}
\bibinfo{author}{\bibfnamefont{C.~W.~J.} \bibnamefont{Beenakker}},
  \bibinfo{journal}{Physical Review Letters} \textbf{\bibinfo{volume}{97}}
  (\bibinfo{year}{2006}).

\bibitem[{\citenamefont{Beenakker}(2008)}]{SAR2}
\bibinfo{author}{\bibfnamefont{C.~W.~J.} \bibnamefont{Beenakker}},
  \bibinfo{journal}{Reviews of Modern Physics} \textbf{\bibinfo{volume}{80}},
  \bibinfo{pages}{1337} (\bibinfo{year}{2008}).

\bibitem[{\citenamefont{Klapwijk}(2004)}]{AR2}
\bibinfo{author}{\bibfnamefont{T.~M.} \bibnamefont{Klapwijk}},
  \bibinfo{journal}{Journal of Superconductivity}
  \textbf{\bibinfo{volume}{17}}, \bibinfo{pages}{593} (\bibinfo{year}{2004}).

\bibitem[{\citenamefont{Pannetier and Courtois}(2000)}]{AR3}
\bibinfo{author}{\bibfnamefont{B.}~\bibnamefont{Pannetier}} \bibnamefont{and}
  \bibinfo{author}{\bibfnamefont{H.}~\bibnamefont{Courtois}},
  \bibinfo{journal}{Journal of Low Temperature Physics}
  \textbf{\bibinfo{volume}{118}}, \bibinfo{pages}{599} (\bibinfo{year}{2000}).

\bibitem[{\citenamefont{Blonder et~al.}(1982)\citenamefont{Blonder, Tinkham,
  and Klapwijk}}]{BKT1}
\bibinfo{author}{\bibfnamefont{G.~E.} \bibnamefont{Blonder}},
  \bibinfo{author}{\bibfnamefont{M.}~\bibnamefont{Tinkham}}, \bibnamefont{and}
  \bibinfo{author}{\bibfnamefont{T.~M.} \bibnamefont{Klapwijk}},
  \bibinfo{journal}{Physical Review B} \textbf{\bibinfo{volume}{25}},
  \bibinfo{pages}{4515} (\bibinfo{year}{1982}).

\bibitem[{\citenamefont{Das~Sarma et~al.}(2011)\citenamefont{Das~Sarma, Adam,
  Hwang, and Rossi}}]{graphene}
\bibinfo{author}{\bibfnamefont{S.}~\bibnamefont{Das~Sarma}},
  \bibinfo{author}{\bibfnamefont{S.}~\bibnamefont{Adam}},
  \bibinfo{author}{\bibfnamefont{E.~H.} \bibnamefont{Hwang}}, \bibnamefont{and}
  \bibinfo{author}{\bibfnamefont{E.}~\bibnamefont{Rossi}},
  \bibinfo{journal}{Reviews of Modern Physics} \textbf{\bibinfo{volume}{83}},
  \bibinfo{pages}{407} (\bibinfo{year}{2011}).

\bibitem[{\citenamefont{Wallace}(1947)}]{BLG1}
\bibinfo{author}{\bibfnamefont{P.~R.} \bibnamefont{Wallace}},
  \bibinfo{journal}{Physical Review} \textbf{\bibinfo{volume}{71}},
  \bibinfo{pages}{622} (\bibinfo{year}{1947}).

\bibitem[{\citenamefont{McCann and Fal'ko}(2006)}]{BLG2}
\bibinfo{author}{\bibfnamefont{E.}~\bibnamefont{McCann}} \bibnamefont{and}
  \bibinfo{author}{\bibfnamefont{V.}~\bibnamefont{Fal'ko}},
  \bibinfo{journal}{Physical Review Letters} \textbf{\bibinfo{volume}{96}},
  \bibinfo{pages}{086805} (\bibinfo{year}{2006}).

\bibitem[{\citenamefont{Martin et~al.}(2008)\citenamefont{Martin, Akerman,
  Ulbricht, Lohmann, Smet, Von~Klitzing, and Yacoby}}]{puddles1}
\bibinfo{author}{\bibfnamefont{J.}~\bibnamefont{Martin}},
  \bibinfo{author}{\bibfnamefont{N.}~\bibnamefont{Akerman}},
  \bibinfo{author}{\bibfnamefont{G.}~\bibnamefont{Ulbricht}},
  \bibinfo{author}{\bibfnamefont{T.}~\bibnamefont{Lohmann}},
  \bibinfo{author}{\bibfnamefont{J.~H.} \bibnamefont{Smet}},
  \bibinfo{author}{\bibfnamefont{K.}~\bibnamefont{Von~Klitzing}},
  \bibnamefont{and} \bibinfo{author}{\bibfnamefont{A.}~\bibnamefont{Yacoby}},
  \bibinfo{journal}{Nature Physics} \textbf{\bibinfo{volume}{4}},
  \bibinfo{pages}{144} (\bibinfo{year}{2008}).

\bibitem[{\citenamefont{Bolotin et~al.}(2008)\citenamefont{Bolotin, Sikes,
  Hone, Stormer, and Kim}}]{susp}
\bibinfo{author}{\bibfnamefont{K.}~\bibnamefont{Bolotin}},
  \bibinfo{author}{\bibfnamefont{K.}~\bibnamefont{Sikes}},
  \bibinfo{author}{\bibfnamefont{J.}~\bibnamefont{Hone}},
  \bibinfo{author}{\bibfnamefont{H.}~\bibnamefont{Stormer}}, \bibnamefont{and}
  \bibinfo{author}{\bibfnamefont{P.}~\bibnamefont{Kim}},
  \bibinfo{journal}{Physical Review Letters} \textbf{\bibinfo{volume}{101}},
  \bibinfo{pages}{096802} (\bibinfo{year}{2008}).

\bibitem[{\citenamefont{Dean et~al.}(2010)\citenamefont{Dean, Young, Meric,
  Lee, Wang, Sorgenfrei, Watanabe, Taniguchi, Kim, Shepard et~al.}}]{BN1}
\bibinfo{author}{\bibfnamefont{C.~R.} \bibnamefont{Dean}},
  \bibinfo{author}{\bibfnamefont{A.~F.} \bibnamefont{Young}},
  \bibinfo{author}{\bibfnamefont{I.}~\bibnamefont{Meric}},
  \bibinfo{author}{\bibfnamefont{C.}~\bibnamefont{Lee}},
  \bibinfo{author}{\bibfnamefont{L.}~\bibnamefont{Wang}},
  \bibinfo{author}{\bibfnamefont{S.}~\bibnamefont{Sorgenfrei}},
  \bibinfo{author}{\bibfnamefont{K.}~\bibnamefont{Watanabe}},
  \bibinfo{author}{\bibfnamefont{T.}~\bibnamefont{Taniguchi}},
  \bibinfo{author}{\bibfnamefont{P.}~\bibnamefont{Kim}},
  \bibinfo{author}{\bibfnamefont{K.~L.} \bibnamefont{Shepard}},
  \bibnamefont{et~al.}, \bibinfo{journal}{Nat Nanotechnol}
  \textbf{\bibinfo{volume}{5}}, \bibinfo{pages}{722} (\bibinfo{year}{2010}).

\bibitem[{\citenamefont{Xue et~al.}(2011)\citenamefont{Xue, Sanchez-Yamagishi,
  Bulmash, Jacquod, Deshpande, Watanabe, Taniguchi, Jarillo-Herrero, and
  Leroy}}]{puddles2}
\bibinfo{author}{\bibfnamefont{J.~M.} \bibnamefont{Xue}},
  \bibinfo{author}{\bibfnamefont{J.}~\bibnamefont{Sanchez-Yamagishi}},
  \bibinfo{author}{\bibfnamefont{D.}~\bibnamefont{Bulmash}},
  \bibinfo{author}{\bibfnamefont{P.}~\bibnamefont{Jacquod}},
  \bibinfo{author}{\bibfnamefont{A.}~\bibnamefont{Deshpande}},
  \bibinfo{author}{\bibfnamefont{K.}~\bibnamefont{Watanabe}},
  \bibinfo{author}{\bibfnamefont{T.}~\bibnamefont{Taniguchi}},
  \bibinfo{author}{\bibfnamefont{P.}~\bibnamefont{Jarillo-Herrero}},
  \bibnamefont{and} \bibinfo{author}{\bibfnamefont{B.~J.} \bibnamefont{Leroy}},
  \bibinfo{journal}{Nature Materials} \textbf{\bibinfo{volume}{10}},
  \bibinfo{pages}{282} (\bibinfo{year}{2011}).

\bibitem[{\citenamefont{Heersche et~al.}(2007)\citenamefont{Heersche,
  Jarillo-Herrero, Oostinga, Vandersypen, and Morpurgo}}]{JEG1}
\bibinfo{author}{\bibfnamefont{H.~B.} \bibnamefont{Heersche}},
  \bibinfo{author}{\bibfnamefont{P.}~\bibnamefont{Jarillo-Herrero}},
  \bibinfo{author}{\bibfnamefont{J.~B.} \bibnamefont{Oostinga}},
  \bibinfo{author}{\bibfnamefont{L.~M.~K.} \bibnamefont{Vandersypen}},
  \bibnamefont{and} \bibinfo{author}{\bibfnamefont{A.~F.}
  \bibnamefont{Morpurgo}}, \bibinfo{journal}{Nature}
  \textbf{\bibinfo{volume}{446}}, \bibinfo{pages}{56} (\bibinfo{year}{2007}).

\bibitem[{\citenamefont{Du et~al.}(2008)\citenamefont{Du, Skachko, and
  Andrei}}]{JEG2}
\bibinfo{author}{\bibfnamefont{X.}~\bibnamefont{Du}},
  \bibinfo{author}{\bibfnamefont{I.}~\bibnamefont{Skachko}}, \bibnamefont{and}
  \bibinfo{author}{\bibfnamefont{E.~Y.} \bibnamefont{Andrei}},
  \bibinfo{journal}{Physical Review B} \textbf{\bibinfo{volume}{77}}
  (\bibinfo{year}{2008}).

\bibitem[{\citenamefont{Miao et~al.}(2007)\citenamefont{Miao, Wijeratne, Zhang,
  Coskun, Bao, and Lau}}]{JEG3}
\bibinfo{author}{\bibfnamefont{F.}~\bibnamefont{Miao}},
  \bibinfo{author}{\bibfnamefont{S.}~\bibnamefont{Wijeratne}},
  \bibinfo{author}{\bibfnamefont{Y.}~\bibnamefont{Zhang}},
  \bibinfo{author}{\bibfnamefont{U.~C.} \bibnamefont{Coskun}},
  \bibinfo{author}{\bibfnamefont{W.}~\bibnamefont{Bao}}, \bibnamefont{and}
  \bibinfo{author}{\bibfnamefont{C.~N.} \bibnamefont{Lau}},
  \bibinfo{journal}{Science} \textbf{\bibinfo{volume}{317}},
  \bibinfo{pages}{1530} (\bibinfo{year}{2007}).

\bibitem[{\citenamefont{Popinciuc et~al.}(2012)\citenamefont{Popinciuc, Calado,
  Liu, Akhmerov, Klapwijk, and Vandersypen}}]{JEG4}
\bibinfo{author}{\bibfnamefont{M.}~\bibnamefont{Popinciuc}},
  \bibinfo{author}{\bibfnamefont{V.~E.} \bibnamefont{Calado}},
  \bibinfo{author}{\bibfnamefont{X.~L.} \bibnamefont{Liu}},
  \bibinfo{author}{\bibfnamefont{A.~R.} \bibnamefont{Akhmerov}},
  \bibinfo{author}{\bibfnamefont{T.~M.} \bibnamefont{Klapwijk}},
  \bibnamefont{and} \bibinfo{author}{\bibfnamefont{L.~M.~K.}
  \bibnamefont{Vandersypen}}, \bibinfo{journal}{Physical Review B}
  \textbf{\bibinfo{volume}{85}} (\bibinfo{year}{2012}).

\bibitem[{\citenamefont{Rickhaus et~al.}(2012)\citenamefont{Rickhaus, Weiss,
  Marot, and Schonenberger}}]{JEG5}
\bibinfo{author}{\bibfnamefont{P.}~\bibnamefont{Rickhaus}},
  \bibinfo{author}{\bibfnamefont{M.}~\bibnamefont{Weiss}},
  \bibinfo{author}{\bibfnamefont{L.}~\bibnamefont{Marot}}, \bibnamefont{and}
  \bibinfo{author}{\bibfnamefont{C.}~\bibnamefont{Schonenberger}},
  \bibinfo{journal}{Nano Letters} \textbf{\bibinfo{volume}{12}},
  \bibinfo{pages}{1942} (\bibinfo{year}{2012}).

\bibitem[{\citenamefont{Calado et~al.}(2015)\citenamefont{Calado, Goswami,
  Nanda, Diez, Akhmerov, Watanabe, Taniguchi, Klapwijk, and
  Vandersypen}}]{Delft}
\bibinfo{author}{\bibfnamefont{V.~E.} \bibnamefont{Calado}},
  \bibinfo{author}{\bibfnamefont{S.}~\bibnamefont{Goswami}},
  \bibinfo{author}{\bibfnamefont{G.}~\bibnamefont{Nanda}},
  \bibinfo{author}{\bibfnamefont{M.}~\bibnamefont{Diez}},
  \bibinfo{author}{\bibfnamefont{A.}~\bibnamefont{Akhmerov}},
  \bibinfo{author}{\bibfnamefont{K.}~\bibnamefont{Watanabe}},
  \bibinfo{author}{\bibfnamefont{T.}~\bibnamefont{Taniguchi}},
  \bibinfo{author}{\bibfnamefont{T.}~\bibnamefont{Klapwijk}}, \bibnamefont{and}
  \bibinfo{author}{\bibfnamefont{L.}~\bibnamefont{Vandersypen}},
  \bibinfo{journal}{ArXiv} p. \bibinfo{pages}{1501.06817}
  (\bibinfo{year}{2015}).

\bibitem[{\citenamefont{Ben~Shalom et~al.}(2015)\citenamefont{Ben~Shalom, Zhu,
  Fal'ko, Mishchenko, Kretinin, Novoselov, Woods, Watanabe, Taniguchi, Geim
  et~al.}}]{Geim}
\bibinfo{author}{\bibfnamefont{M.}~\bibnamefont{Ben~Shalom}},
  \bibinfo{author}{\bibfnamefont{M.~J.} \bibnamefont{Zhu}},
  \bibinfo{author}{\bibfnamefont{V.~I.} \bibnamefont{Fal'ko}},
  \bibinfo{author}{\bibfnamefont{A.}~\bibnamefont{Mishchenko}},
  \bibinfo{author}{\bibfnamefont{A.~V.} \bibnamefont{Kretinin}},
  \bibinfo{author}{\bibfnamefont{K.~S.} \bibnamefont{Novoselov}},
  \bibinfo{author}{\bibfnamefont{C.~R.} \bibnamefont{Woods}},
  \bibinfo{author}{\bibfnamefont{K.}~\bibnamefont{Watanabe}},
  \bibinfo{author}{\bibfnamefont{T.}~\bibnamefont{Taniguchi}},
  \bibinfo{author}{\bibfnamefont{A.~K.} \bibnamefont{Geim}},
  \bibnamefont{et~al.}, \bibinfo{journal}{ArXiv} p. \bibinfo{pages}{1504.03286}
  (\bibinfo{year}{2015}).

\bibitem[{\citenamefont{Clayman and Frindt}(1971)}]{NbSe1}
\bibinfo{author}{\bibfnamefont{B.~P.} \bibnamefont{Clayman}} \bibnamefont{and}
  \bibinfo{author}{\bibfnamefont{R.~F.} \bibnamefont{Frindt}},
  \bibinfo{journal}{Solid State Communications} \textbf{\bibinfo{volume}{9}},
  \bibinfo{pages}{1881} (\bibinfo{year}{1971}).

\bibitem[{\citenamefont{Frindt}(1972)}]{NbSe2}
\bibinfo{author}{\bibfnamefont{R.~F.} \bibnamefont{Frindt}},
  \bibinfo{journal}{Physical Review Letters} \textbf{\bibinfo{volume}{28}},
  \bibinfo{pages}{299} (\bibinfo{year}{1972}).

\bibitem[{\citenamefont{Staley et~al.}(2009)\citenamefont{Staley, Wu, Eklund,
  Liu, Li, and Xu}}]{NbSe3}
\bibinfo{author}{\bibfnamefont{N.~E.} \bibnamefont{Staley}},
  \bibinfo{author}{\bibfnamefont{J.}~\bibnamefont{Wu}},
  \bibinfo{author}{\bibfnamefont{P.}~\bibnamefont{Eklund}},
  \bibinfo{author}{\bibfnamefont{Y.}~\bibnamefont{Liu}},
  \bibinfo{author}{\bibfnamefont{L.~J.} \bibnamefont{Li}}, \bibnamefont{and}
  \bibinfo{author}{\bibfnamefont{Z.}~\bibnamefont{Xu}},
  \bibinfo{journal}{Physical Review B} \textbf{\bibinfo{volume}{80}}
  (\bibinfo{year}{2009}).

\bibitem[{\citenamefont{Geim and Grigorieva}(2013)}]{vdW1}
\bibinfo{author}{\bibfnamefont{A.~K.} \bibnamefont{Geim}} \bibnamefont{and}
  \bibinfo{author}{\bibfnamefont{I.~V.} \bibnamefont{Grigorieva}},
  \bibinfo{journal}{Nature} \textbf{\bibinfo{volume}{499}},
  \bibinfo{pages}{419} (\bibinfo{year}{2013}).

\bibitem[{xxx()}]{xxx}
\emph{\bibinfo{title}{The indirect estimates for $\delta \epsilon_{F}$ from
  measurements of the FWHM of the $R_{xx}$-peak are typically found to be much
  higher than form the direct measurement of $\delta \epsilon_{F}$ by scanning
  gate or scanning tunneling microscopy.}} (????).

\bibitem[{\citenamefont{Ludwig}(2007)}]{SAR3}
\bibinfo{author}{\bibfnamefont{T.}~\bibnamefont{Ludwig}},
  \bibinfo{journal}{Physical Review B} \textbf{\bibinfo{volume}{75}}
  (\bibinfo{year}{2007}).

\bibitem[{\citenamefont{Blake et~al.}(2009)\citenamefont{Blake, Yang, Morozov,
  Schedin, Ponomarenko, Zhukov, Nair, Grigorieva, Novoselov, and Geim}}]{cnt3}
\bibinfo{author}{\bibfnamefont{P.}~\bibnamefont{Blake}},
  \bibinfo{author}{\bibfnamefont{R.}~\bibnamefont{Yang}},
  \bibinfo{author}{\bibfnamefont{S.~V.} \bibnamefont{Morozov}},
  \bibinfo{author}{\bibfnamefont{F.}~\bibnamefont{Schedin}},
  \bibinfo{author}{\bibfnamefont{L.~A.} \bibnamefont{Ponomarenko}},
  \bibinfo{author}{\bibfnamefont{A.~A.} \bibnamefont{Zhukov}},
  \bibinfo{author}{\bibfnamefont{R.~R.} \bibnamefont{Nair}},
  \bibinfo{author}{\bibfnamefont{I.~V.} \bibnamefont{Grigorieva}},
  \bibinfo{author}{\bibfnamefont{K.~S.} \bibnamefont{Novoselov}},
  \bibnamefont{and} \bibinfo{author}{\bibfnamefont{A.~K.} \bibnamefont{Geim}},
  \bibinfo{journal}{Solid State Communications} \textbf{\bibinfo{volume}{149}},
  \bibinfo{pages}{1068 } (\bibinfo{year}{2009}).

\bibitem[{\citenamefont{Xia et~al.}(2011)\citenamefont{Xia, Perebeinos, Lin,
  Wu, and Avouris}}]{cnt1}
\bibinfo{author}{\bibfnamefont{F.~N.} \bibnamefont{Xia}},
  \bibinfo{author}{\bibfnamefont{V.}~\bibnamefont{Perebeinos}},
  \bibinfo{author}{\bibfnamefont{Y.~M.} \bibnamefont{Lin}},
  \bibinfo{author}{\bibfnamefont{Y.~Q.} \bibnamefont{Wu}}, \bibnamefont{and}
  \bibinfo{author}{\bibfnamefont{P.}~\bibnamefont{Avouris}},
  \bibinfo{journal}{Nature Nanotechnology} \textbf{\bibinfo{volume}{6}}
  (\bibinfo{year}{2011}).

\bibitem[{\citenamefont{Wang et~al.}(2013)\citenamefont{Wang, Meric, Huang,
  Gao, Gao, Tran, Taniguchi, Watanabe, Campos, Muller et~al.}}]{cnt2}
\bibinfo{author}{\bibfnamefont{L.}~\bibnamefont{Wang}},
  \bibinfo{author}{\bibfnamefont{I.}~\bibnamefont{Meric}},
  \bibinfo{author}{\bibfnamefont{P.~Y.} \bibnamefont{Huang}},
  \bibinfo{author}{\bibfnamefont{Q.}~\bibnamefont{Gao}},
  \bibinfo{author}{\bibfnamefont{Y.}~\bibnamefont{Gao}},
  \bibinfo{author}{\bibfnamefont{H.}~\bibnamefont{Tran}},
  \bibinfo{author}{\bibfnamefont{T.}~\bibnamefont{Taniguchi}},
  \bibinfo{author}{\bibfnamefont{K.}~\bibnamefont{Watanabe}},
  \bibinfo{author}{\bibfnamefont{L.~M.} \bibnamefont{Campos}},
  \bibinfo{author}{\bibfnamefont{D.~A.} \bibnamefont{Muller}},
  \bibnamefont{et~al.}, \bibinfo{journal}{Science}
  \textbf{\bibinfo{volume}{342}} (\bibinfo{year}{2013}).

\bibitem[{\citenamefont{Nilsson et~al.}(2006)\citenamefont{Nilsson,
  Castro~Neto, Peres, and Guinea}}]{BLG3}
\bibinfo{author}{\bibfnamefont{J.}~\bibnamefont{Nilsson}},
  \bibinfo{author}{\bibfnamefont{A.}~\bibnamefont{Castro~Neto}},
  \bibinfo{author}{\bibfnamefont{N.}~\bibnamefont{Peres}}, \bibnamefont{and}
  \bibinfo{author}{\bibfnamefont{F.}~\bibnamefont{Guinea}},
  \bibinfo{journal}{Physical Review B} \textbf{\bibinfo{volume}{73}},
  \bibinfo{pages}{214418} (\bibinfo{year}{2006}).

\bibitem[{\citenamefont{Beenakker}(1997)}]{RMT}
\bibinfo{author}{\bibfnamefont{C.}~\bibnamefont{Beenakker}},
  \bibinfo{journal}{Reviews of Modern Physics} \textbf{\bibinfo{volume}{69}},
  \bibinfo{pages}{731} (\bibinfo{year}{1997}).

\bibitem[{\citenamefont{Landau and Lifshitz}(1977)}]{LL}
\bibinfo{editor}{\bibfnamefont{L.~D.} \bibnamefont{Landau}} \bibnamefont{and}
  \bibinfo{editor}{\bibfnamefont{E.~M.} \bibnamefont{Lifshitz}}, eds.,
  \emph{\bibinfo{title}{Quantum Mechanics}} (\bibinfo{publisher}{Pergamon},
  \bibinfo{year}{1977}).

\end{thebibliography}

\pagebreak
\begin{center}
\textbf{\large Supplementary Information}
\end{center}
\setcounter{equation}{0}
\setcounter{figure}{0}
\setcounter{table}{0}
\setcounter{page}{1}
\makeatletter
\renewcommand{\theequation}{\arabic{equation}}
\renewcommand{\thefigure}{S\arabic{figure}}
\renewcommand{\bibnumfmt}[1]{[#1]}
\renewcommand{\citenumfont}[1]{#1}

\maketitle
\section{Ohmic electronic interfaces in hBN/bilayer graphene/NbSe$_{2}$ heterostructures}

\begin{figure*}[tbp]
\centering
\includegraphics[width=1\textwidth]{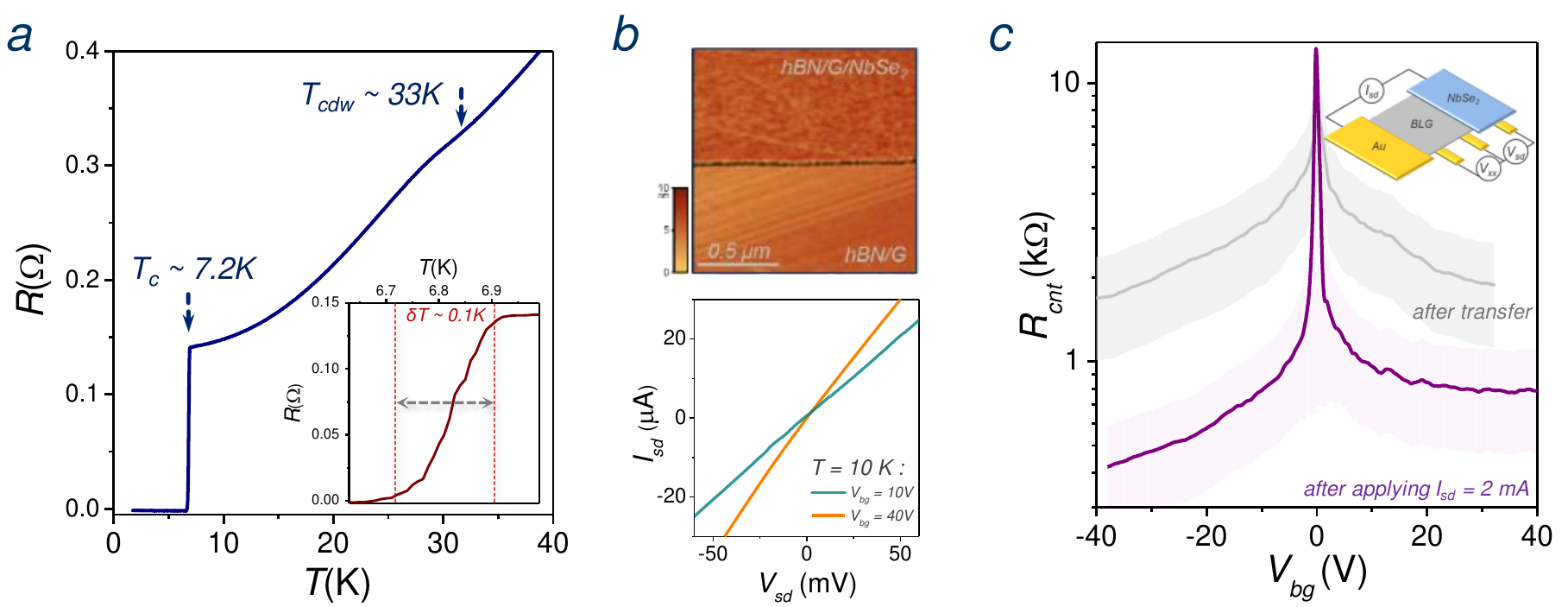}
\caption[Ohmic electronic interfaces in hBN/bilayer graphene/NbSe$_{2}$ hetero-structures]
{(a) Resistance $R$ vs. temperature $T$ of NbSe$_{2}$. The sample undergoes a CDW phase transition below $T_{cdw} \sim 33$~K and becomes SC below NbSe$_{2}$'s $T_{c} \sim 6.8$~K. Both, the ``hump''-like feature in $R$ at the CDW transition and the sharpness of the SC transition, with a width of $\delta T \sim 0.1$~K (inset), demonstrate the high purity of the used NbSe$_{2}$ crystal.
(b) Top : atomic force microscopy image of the atomically flat BLG/NbSe$_{2}$ interfacial region forming a sharp N/SC junction. Bottom : typical current-biased four-terminal $I$-$V$ measurements across the BLG/NbSe$_{2}$ junction for $T > T_{c}$. For all $V_{bg}$ the $I$-$V$ traces are linear, hence showing Ohmic characteristics.
(c) Extracted contact resistance $R_{cnt}$ of the BLG/NbSe$_{2}$ interface as a function of $V_{bg}$ (with error bars), before (red line) and after (grey) current annealing of the junction with $I_{anneal}=2$~mA. The current annealing effectively reduces $R_{cnt}$ at the same time sharpening the region around the CNP. Inset : schematics of the measurement setup.}
\end{figure*}

To fabricate the hBN/BLG/NbSe$_{2}$ hetero-structures we pre-fabricate hBN/BLG Hall-bar devices following the standard recipe in~\cite{BN1}. We intentionally do not contact one end of the patterned BLG with a Ti/Au electrode, leaving it free for the deposition of NbSe$_{2}$. The NbSe$_{2}$ flake is then transferred onto the device with the vdW transfer technique creating an overlap region with the BLG from one side, and four gold electrodes from the other side, hence allowing to inject current from NbSe$_{2}$ directly into BLG. The so created gold/NbSe$_{2}$ and BLG/NbSe$_{2}$ junctions have Ohmic interfacial resistances with typical values of $50$-$500$~$\Omega \mu$m and $1$-$10$~k$\Omega \mu$m respectively (Fig. S1 (b)). Fig. S1 (a) demonstrates typical four-point resistance $R$ vs. temperature $T$ measurements of the NbSe$_{2}$ flake.

We estimate the contact resistance $R_{cnt}$ between BLG and NbSe$_{2}$ with the typical four-probe method~\cite{cnt3}. By injecting a source-drain current $I_{sd}$=$10$~nA across the interface we measure the resulting voltage difference $V_{sd}$ across the device (Fig. 1 (b) and inset Fig. S1 (c)). As one of the voltage probes connects to the pristine NbSe$_{2}$ and the other to the pristine BLG channel, the so measured $V_{sd}$ signal, in all generality, contains contributions from all three regions - the pristine NbSe$_{2}$, the BLG/NbSe$_{2}$ interface and the pristine BLG. By neglecting NbSe$_{2}$ intrinsic resistivity (as it is in the m$\Omega$ range (Fig. S1 (a))) we can estimate the interfacial contact resistance through the relation $R_{cnt}=(V_{sd}-V_{xx})/I_{sd}$, where we correct for BLGs intrinsic resistivity by measuring $V_{xx}$ inside the BLG channel separately and subtracting it from $V_{sd}$. Overall the so obtained strongly back gate dependent $R_{cnt}(V_{bg})$ (Fig. S1 (c)) can be also directly used for a more accurate estimate of the voltage drop across the BLG/NbSe$_{2}$ interface $V_{ns}(V_{bg})=R_{cnt}(V_{bg})I_{sd}$.

Here $R_{cnt}(V_{bg})$ has values as low as $400 \pm 200$~$\Omega \mu m$ (after current annealing) at negative $V_{bg}$ and has asymmetrically higher values at positive $V_{bg}$. Overall this behaviour is in good agreement with typical metal/graphene junctions~\cite{cnt1,cnt2,cnt3} and can be explained by $pn$-junction formation at the junction edge due to work-function matching between the NbSe$_{2}$ and the BLG, with the direct consequence that the majority of the current is injected into the BLG at the contact edge and not at the contact interface~\cite{cnt1}. Here the work-functions $\Phi_{NbSe_{2}}=$~5.9 eV of NbSe$_{2}$ and that of the typically used gold $\Phi_{Au}=$~4.4 eV~\cite{NbSe3} have comparable values.

\section{Theoretical conductance of the BLG/SC interface}

\subsection{General formulas}

We model the experimental setup by a BLG in the normal state in
the half plane $x>0$ having an induced SC gap $\Delta _{0}$ at $%
x < 0$. The proximity to the SC results not only in the generation of
the gap but also changes the chemical potential at $x<0.$

The Bogoliubov-deGennes (BdG) equations describing the electron motion in the
system can be written as~\cite{BdG} :%

\begin{equation}
\scalebox{0.8}{$
\left(
\begin{array}{cc}
\mathcal{H}\left( \mathbf{\hat{K}}\right) +U\left( x\right) - \varepsilon_{F} & \Delta
\left( x\right) \\
\Delta ^{\ast }\left( x\right) & \varepsilon_{F}-\mathcal{H}\left( \mathbf{\hat{K}}%
\right) -U\left( x\right)%
\end{array}%
\right) \left(
\begin{array}{c}
u \\
v%
\end{array}%
\right) =\varepsilon \left(
\begin{array}{c}
u \\
v%
\end{array}%
\right)  \label{0a1} $}
\end{equation}%

where $\mathbf{\hat{K}}$ is the momentum operator and $u$ and $v$ the electron and hole eigenfunctions. Assuming that only the
singlet SC is induced and that only electrons of different
valleys form Cooper pairs we write the Hamiltonian $\mathcal{H}\left(
\mathbf{K}\right) $ of the normal BLG for one valley as a $4\times 4$
matrix~\cite{BLG1,BLG2,BLG3} :

\begin{equation}
\mathcal{H}\left( \mathbf{K}\right) \mathcal{=}\hslash v\left(
\begin{array}{cccc}
0 & Ke^{i\alpha } & -t_{\perp } & 0 \\
Ke^{-i\alpha } & 0 & 0 & 0 \\
-t_{\perp } & 0 & 0 & Ke^{-i\alpha } \\
0 & 0 & Ke^{ia} & 0%
\end{array}%
\right)  \label{0a2}
\end{equation}%

where $\alpha \left( \mathbf{K}\right) =\arctan \left(k_{y}/k_{x}\right) $. Here $k_{y}$ is
component of the momentum $\mathbf{K}$ parallel to the interface, $k_{x}$ is the $%
x$-component of $\mathbf{K}$ and $v$ the Fermi velocity.

Due to the SC and the SC gap $\Delta \left( x\right) $ the potential $U\left( x\right) $ is only finite at $x < 0$ and these quantities
can be written in the form :%

\begin{equation}
U=\left\{
\begin{array}{cc}
-U_{0}, & x<0 \\
0, & x>0%
\end{array}%
\right. \label{0a3}
\end{equation}

\begin{equation}
\Delta =\left\{
\begin{array}{cc}
\Delta _{0}e^{i\phi }, & x<0 \\
0, & x>0%
\end{array}%
\right.  \label{0a4}
\end{equation}%

The conductance of a SLG/SC junction has first been
considered by C.W.J. Beenakker~\cite{SAR1,SAR2}, while the same
quantity was calculated for BLG in a subsequent publication in by T.
Ludwig~\cite{SAR3}. However, in the latter work
only the limit of $\hslash vt_{\perp } \ll U_{0}$ was considered. All this does not allow to make a
detailed comparison of that theory with the present experimental data.

The sharp change of the potential $U\left( x\right) $ and the
SC gap $\Delta \left( x\right) $ at $x=0$ is an
oversimplification. In the experiment the dependence of these quantities on
the coordinates can be rather smooth. Although this may result in a certain
overestimation of the transmission amplitude at the interface, we do not
think that the dependence of the conductance on $\varepsilon_{F}$ and $eV_{ns}$ can
essentially be different in a model with a more smooth potential $U\left(
x\right) $.

Calculating the conductance one considers scattering of particles having the
energy $\varepsilon = eV_{ns}$ (counted from $\varepsilon_{F}$) and moving from right to the left.
Here, in accordance with the experiment we assume that $U_{0}\gg \varepsilon_{F},\Delta _{0},eV_{ns}$, with the largest energy
in being the coupling energy $\hslash vt_{\perp }$ between the layers $\hslash vt_{\perp }\gg U_{0}$.
As the experimental measurements have been presented for $e\left\vert
V_{ns}\right\vert <\Delta _{0},$ we calculate the differential conductance $%
G_{ns}\left( \varepsilon \right) $ for $\left\vert \varepsilon \right\vert
<\Delta _{0}$. In this region, using a formalism developed by C.J.W.
Beenakker\cite{RMT}, one can reduce the
calculation of the conductance $G_{ns}\left( \varepsilon \right) $ of a
N/SC interface to calculation of the transmission $%
t\left( \varepsilon \right) $ and reflection amplitudes $r\left( \varepsilon
\right) $ for the same interface but in the absence of the SC gap.

According to this theory the differential conductance $G_{ns}\left(\varepsilon \right) $ takes the form :%

\begin{equation}
G_{ns}\left( \varepsilon \right) =2G_{0}Tr\left[ m\left( \varepsilon\right)
m^{+}\left( \varepsilon\right) \right]   \label{0a7}
\end{equation}%

where $G_{0}=2e^{2}/h$ and :%

\begin{equation}
m\left( \varepsilon \right) =t_{12}\left( \varepsilon \right) \left[
1-e^{-2i\beta }r_{22}^{\ast }\left( -\varepsilon \right) r_{22}\left(
\varepsilon \right) \right] ^{-1}t_{21}^{\ast }\left( -\varepsilon \right)
\label{0a8}
\end{equation}%

with :%

\begin{equation}
\beta =\arccos \left( \varepsilon /\Delta _{0}\right)  \label{0a9}
\end{equation}%

Eq. (\ref{0a7}-\ref{0a9}) show that calculating the transmission $%
t_{12}\left( \varepsilon \right) $, $t_{21}\left( \varepsilon \right) ,$ and
reflection $r_{12}\left( \varepsilon \right) ,$ $r_{21}\left( \varepsilon
\right) $ amplitudes for right and left moving particles at the interface
between normal metals one obtains the differential conductance $G_{ns}\left(
\varepsilon \right) $ of the N/SC interface. The trace $Tr
$ over the scattering channels in Eq. (\ref{0a7}) reduces in the present
model to integration over $k_{y}$. Overall Eqs. (\ref{0a7}-\ref{0a9}) demonstrate that in order to calculate the
differential conductance $G_{ns}$, one should simply understand the scattering
on the interface between two normal metals.

Experimentally, it is convenient to plot not the conductance $G_{ns}\left(
\varepsilon \right) $ itself but the ratio $G_{ns}\left( \varepsilon \right)
/G_{nn}\left( \varepsilon \right)$, where $G_{nn}\left( \varepsilon \right)
$ is the conductance of the interface at temperatures exceeding the
SC transition temperature $T_{c}$. $G_{nn}\left( \varepsilon ,\varepsilon_{F}\right) $ can be easily expressed in terms
of the transmission coefficient as :

\begin{equation}
G_{nn}\left( \varepsilon ,\varepsilon_{F}\right) =2G_{0}Tr\left\vert t_{12}\left(
\varepsilon ,\varepsilon_{F}\right) \right\vert ^{2}  \label{0a10}
\end{equation}%

At the same time, the conductance $G_{nn}\left( \varepsilon ,\varepsilon_{F}\right) $
given by Eq. (\ref{0a10}) is applicable at low temperatures only. In the
present experiment the temperature at which the normal conductance
is measured exceeds the energies $\varepsilon_{F}$ and $\Delta _{0}$. In this case one
should take a thermodynamic average $\left\langle G_{nn}\left( \varepsilon
,\varepsilon_{F}\right) \right\rangle _{F}$ using the standard formula :%

\begin{equation}
\left\langle G_{nn}\left( \varepsilon ,\varepsilon_{F},T\right) \right\rangle _{F}=%
\frac{1}{4T}\int_{-\infty }^{\infty }G_{nn}\left( \varepsilon ,\mu \right)
\frac{d\mu }{\cosh ^{2}\left( \frac{\mu -\varepsilon_{F}}{2T}\right) }  \label{0a11}
\end{equation}%

Eq. (\ref{0a11}) is used in the present work for the calculation of the physical
conductance at $T > T_{c}$. At the same time, the measurements below $T_{c}$
have been performed at sufficiently low temperature, such that one can use
the formulas derived at $T=0.$

One can see from Eqs. (\ref{0a7}, \ref{0a8}) an important difference between
the conductance of SLG considered in~\cite{SAR1} and
that of BLG. In the former case the transmission amplitude is of
order one for any $U_{0}$ due to the so called Klein effect and is equal to
a normally incident electron, whereas in the latter case the
transmission amplitude decays proportionally to $U_{0}^{-1/2}$ at large $%
U_{0}$ and can be very small. Then, for BLG the conductance $G_{nn}$
is proportional to $\left\vert t\right\vert ^{2}$, while $G_{ns},$ being
proportional to $\left\vert t\right\vert ^{4}$ can be much smaller than $%
G_{nn}.$

\subsection{Transmission $t\left( \protect\varepsilon \right) $ and
reflection $r\left( \protect\varepsilon \right) $ amplitudes in the absence
of a SC gap.}

\subsubsection{Eigenvalues}

Study of the model with the Hamiltonian $\mathcal{H}\left( \mathbf{K}\right)
$ can give results for an arbitrary relation between $t_{\perp }$ and
characteristic energies inside one layer.

The eigenenergies $E\left( \mathbf{K}\right) $ of the Hamiltonian $\mathcal{H%
}\left( \mathbf{K}\right)$ in Eq. (\ref{0a2}) take the form :%

\begin{eqnarray}
E_{1}\left( \mathbf{K}\right) &=&\hslash v\left( -t_{\perp }/2-\mathcal{E}%
\left( \mathbf{K}\right) \right) ,  \label{k2} \\
E_{2}\left( \mathbf{K}\right) &=&\hslash v\left( -t_{\perp }/2+\mathcal{E}%
\left( \mathbf{K}\right) \right) ,  \notag \\
E_{3}\left( \mathbf{K}\right) &=&\hslash v\left( t_{\perp }/2+\mathcal{E}%
\left( \mathbf{K}\right) \right) ,  \notag \\
E_{4}\left( \mathbf{K}\right) &=&\hslash v\left( t_{\perp }/2-\mathcal{E}%
\left( \mathbf{K}\right) \right)   \notag
\end{eqnarray}%

where : %

\begin{equation}
\mathcal{E}\left( \mathbf{K}\right) =\sqrt{\left( t_{\perp }/2\right)
^{2}+K^{2}}  \label{k3}
\end{equation}

Provided that $t_{\perp }$ exceeds all the other
energies, the bands of the Hamiltonian (\ref{0a2}) with the spectra $%
E_{1}\left( \mathbf{k}\right) $ and $E_{3}\left( \mathbf{k}\right) $ are far
away from the Fermi surface and their contribution into physical quantities
can be neglected. The eigenenergies of the first two low energy bands take
in the limit of small $k_{x} \ll t_{\perp }$ the
following form :%

\begin{eqnarray}
E_{2}\left( k_{x}\right) &=&\hslash v\left( -t_{\perp }/2+\mathcal{E}%
\left( k_{x}\right) \right) \approx \frac{\hslash vK^{2}}{t_{\perp }},
\label{k4} \\
E_{4}\left( k_{x}\right) &=&\hslash v\left( t_{\perp }/2-\mathcal{E}%
\left( k_{x}\right) \right) \approx -\frac{\hslash vK^{2}}{t_{\perp }}
\notag
\end{eqnarray}%

In Eq. (\ref{k4}) $E_{2}\left( \mathbf{K}\right) $ describes the conduction
band (upper part of the spectrum) and $E_{4}\left( \mathbf{K}\right) $
describes the valence band (lower part of the spectrum). Using the
inequality $\hslash vt_{\perp }\gg U_{0}$ we consider only these low lying bands.

\subsubsection{Wave functions in the region $x>0$.}

Starting with the Hamiltonian $\mathcal{H}\left( \mathbf{k}\right) $, Eq. (%
\ref{0a2}), and using a diagonalization procedure of \cite{BLG3}, we
can write out $4$-component vectors $u$ satisfying the equation :%

\begin{equation}
\left( \mathcal{H}\left( k_{x} -\epsilon_{F}\right) u=\varepsilon u \right.
\label{k5a}
\end{equation}%

We consider the case $\varepsilon_{F} > 0$ but the solutions depend on the sign of $%
\varepsilon + \varepsilon_{F}$ and we write them separately for $\varepsilon + \varepsilon_{F} > 0$
and $\varepsilon + \varepsilon_{F} < 0$. For compact notations we introduce the quantity :%

\begin{equation}
K_{0}=\sqrt{\left\vert \varepsilon + \varepsilon_{F}\right\vert t_{\perp }/\hslash v} \label{0a12}
\end{equation}%

We obtain for the low lying eigenvalues :%

\begin{eqnarray}
\varepsilon _{2} &=&-\varepsilon_{F}+\frac{\hslash v}{t_{\perp }}K^{2},\quad
\label{0a21} \\
\varepsilon _{4} &=&-\varepsilon_{F}-\frac{\hslash v}{t_{\perp }}K^{2}  \label{0a19}
\end{eqnarray}%

In order to calculate the wave functions one should choose an eigenvalue $%
\varepsilon $ and determine $K$ as a function of $\varepsilon $. It is
clear that constructing plain waves in the region of the normal metal $N_{1}$
and $\varepsilon_{F} > 0$ one should take the solution (\ref{0a21}) for $K$ at $%
\varepsilon > - \varepsilon_{F}$ and of (\ref{0a19}) at $\varepsilon < -\varepsilon_{F}$. In
addition, one has a solution for $K$ of (\ref{0a19}) at $\varepsilon > - \varepsilon_{F}$
and of (\ref{0a21}) at $\varepsilon < - \varepsilon_{F}$. However, the latter solutions
are exponentially growing or decaying as functions of $x$. Nevertheless,
they should also be taken into account when matching functions at the
interface because of the deep potential $-U_{0}$ at $x < 0$ where the
exponential growth can change to plain wave behavior.

\paragraph{Plain wave solutions at $\protect\varepsilon + \varepsilon_{F} > 0.$}

In this region we have left and right moving electrons from the conduction
band with the energies from Eq. (\ref{0a21}). The solution $%
u_{1}^{R}$ for right moving particles in this region takes the form :%

\begin{equation}
u_{1}^{R}=\frac{e^{ik_{x}x+ik_{y}y}}{2\sqrt{K_{0}/t_{\perp }\cos \alpha }}\left(
\begin{array}{c}
K_{0}/t_{\perp } \\
e^{-i\alpha } \\
K_{0}/t_{\perp } \\
e^{i\alpha }%
\end{array}%
\right) ,\quad  \label{0a13}
\end{equation}%

while the solution for the left moving particles $u_{1}^{L}$ reads :%

\begin{equation}
u_{1}^{L}=\frac{e^{-ik_{x}x+ik_{y}y}}{2\sqrt{K_{0}/t_{\perp }\cos \alpha }}\left(
\begin{array}{c}
-K_{0}/t_{\perp } \\
e^{i\alpha } \\
-K_{0}/t_{\perp } \\
e^{-i\alpha }%
\end{array}%
\right) .  \label{0a14}
\end{equation}%

The wave functions $u_{1}^{R}$ and $u_{2}^{L}$ correspond to the eigenvalue $%
E_{2}\left( \mathbf{k}\right) $ in Eq. (\ref{k4}) and belong to the
conduction band. They are normalized assuming the current $1$ along the $x$%
-axis for right moving particles and $-1$ for left moving ones.

Having fixed $\varepsilon_{F} > 0$ we have to express $k_{x}$ and $k_{y}$ in terms of $K_{0}$
and $\alpha $. Here we introduce the angle $\alpha $ as :%

\begin{equation}
k_{x}-ik_{y}=K_{0}e^{-i\alpha }  \label{0a17}
\end{equation}%

Then, for $\varepsilon + \varepsilon_{F} > 0$ we have the following relations for the
variables $\alpha $ and $k_{x}$, with $\alpha $ varying in the interval $-\pi /2 < \alpha < \pi /2$ :%

\begin{equation}
k_{x}=K_{0}\cos \alpha ,\quad k_{y}=K_{0}\sin \alpha \quad  \label{0a18}
\end{equation}%

\paragraph{Plain wave solutions at $\protect\varepsilon +\varepsilon_{F} < 0$.}

For these energies we have plain waves corresponding to right and
left moving holes from the valence band in Eq. (\ref{0a19}). For the
right moving holes we obtain the normalized wave functions :%

\begin{equation}
u_{2}^{R}=\frac{e^{ik_{x}^{\prime }x+ik_{y}^{\prime}y}}{2\sqrt{K_{0}/t_{\perp }\cos \alpha
^{\prime }}}\left(
\begin{array}{c}
K_{0}/t_{\perp } \\
e^{-i\alpha ^{\prime }} \\
-K_{0}/t_{\perp } \\
-e^{i\alpha ^{\prime }}%
\end{array}%
\right) ,\quad  \label{0a22}
\end{equation}%

where again the angle $\alpha ^{\prime }$ varies in the interval $-\pi /2 < \alpha ^{\prime } < \pi /2.$ :%

\begin{equation}
k_{x}^{\prime }=-K_{0}\cos \alpha ^{\prime },\quad k_{y}^{\prime }=-K_{0}\sin \alpha ^{\prime
}.  \label{0a24}
\end{equation}%

Here, the opposite signs in Eq. (\ref{0a18}) and Eq. (\ref{0a24}) are
due to the fact that we now consider holes from the valence band instead of
electrons from the conduction band. In general the current of the right moving holes equals to $1,$
while the current for the left moving holes equals to $-1 $. The solution $u_{2}^{L}$ for the left moving particles takes the form :%

\begin{equation}
u_{2}^{L}=\frac{e^{-ik_{x}^{\prime }x+ik_{y}^{\prime }y}}{2\sqrt{K_{0}/t_{\perp }\cos \alpha
^{\prime }}}\left(
\begin{array}{c}
-K_{0}/t_{\perp } \\
e^{i\alpha ^{\prime }} \\
K_{0}/t_{\perp } \\
-e^{-i\alpha ^{\prime }}%
\end{array}%
\right) .  \label{0a26}
\end{equation}%

\paragraph{Decaying and growing solutions at $\protect\varepsilon + \varepsilon_{F} < 0.$}

In addition to the plain waves in Eqs. (\ref{0a13}, \ref{0a14}), there are two
other solutions $u_{1}^{<}$ and $u_{1}^{>}$ corresponding to the eigenvalue $%
\varepsilon _{2}$ from the conduction band (here as the normalization does not play any role, we omit the pre-factors) :%

\begin{equation}
u_{1}^{<}=e^{\kappa_{x} x+i\kappa_{y}y}\left(
\begin{array}{c}
-iK_{0}/t_{\perp } \\
e^{\gamma } \\
-iK_{0}/t_{\perp } \\
e^{-\gamma }%
\end{array}%
\right)  \label{0a15}
\end{equation}%

and :%

\begin{equation}
u_{1}^{>}=e^{-\kappa_{x} x+i\kappa_{y}y}\left(
\begin{array}{c}
iK_{0}/t_{\perp } \\
e^{-\gamma } \\
iK_{0}/t_{\perp } \\
e^{\gamma }%
\end{array}%
\right)  \label{0a16}
\end{equation}%

With the parameters $\kappa_{x} $ and $\kappa_{y}$ can be written in the form :%

\begin{equation}
\kappa_{x} =K_{0}\cosh \gamma ,\quad \kappa_{y}=K_{0}\sinh \gamma  \label{0a27}
\end{equation}%

\paragraph{Decaying and growing solutions at $\protect\varepsilon + \varepsilon_{F} > 0.$}

In this region the growing and decaying wave functions
correspond to $\varepsilon _{4}$ in Eq. (\ref{0a19}) describing the
valence band. We write the growing $u_{2}^{<}$ and decaying $u_{2}^{>}$
functions as :%

\begin{equation}
u_{2}^{<}=e^{\kappa_{x}^{\prime } x+i\kappa_{y}^{\prime }y}\left(
\begin{array}{c}
iK_{0}/t_{\perp } \\
e^{\gamma ^{\prime }} \\
-iK_{0}/t_{\perp } \\
-e^{-\gamma ^{\prime }}%
\end{array}%
\right)  \label{0a28}
\end{equation}%

and :%

\begin{equation}
u_{2}^{>}=e^{-\kappa_{x}^{\prime } x+i\kappa_{y}^{\prime }y}\left(
\begin{array}{c}
-iK_{0}/t_{\perp } \\
e^{-\gamma ^{\prime }} \\
iK_{0}/t_{\perp } \\
-e^{\gamma ^{\prime }}%
\end{array}%
\right)  \label{0a29}
\end{equation}%

The parameters $\kappa_{x}^{\prime } $ and $\kappa_{y}^{\prime }$ are :%

\begin{equation}
\kappa_{x}^{\prime } =K_{0}\cosh \gamma ^{\prime },\quad \kappa_{y}^{\prime }=K_{0}\sinh \gamma ^{\prime }
\label{0a30}
\end{equation}%

Overall the wave functions written in the region $x > 0 $ can also be used in the
region $x < 0$ after shifting $\varepsilon_{F}\rightarrow \varepsilon_{F} + U_{0}$. We denote these
functions by adding the subscript $U_{0}$, thus obtaining $u_{1U_{0}}^{R,L},$
$u_{2U_{0}}^{R,L},$ etc.

\subsubsection{Transmission $t_{21}$ and reflection $r_{21}$ amplitudes.}

Now we calculate the transmission $t_{21}\left( \varepsilon \right) $ and
reflection $r_{21}\left( \varepsilon \right) $ amplitudes that match the wave
functions written on the left and on the right of the interface. Again, we
should consider the regions $\varepsilon + \varepsilon_{F} > 0$ and $\varepsilon + \varepsilon_{F} < 0$
separately.

\paragraph{Region $\varepsilon_{F} > 0$, $\protect\varepsilon + \varepsilon_{F} > 0$.}

The scattering process in $N_{2}$ includes a plane wave incident from the
left, $u_{1U_{0}}^{R},$ and another one, $u_{1U_{0}}^{L},$ reflected from
the interface. At the same time, region $N_{2}$ has an additional solution that is
growing with $x$ (decaying from the interface) with the symmetry of $%
u_{1}^{<}$ from Eq. (\ref{0a15}).\qquad

After scattering on the interface of $N_{1}$ one obtains an outgoing wave
with the structure $u_{1}^{R},$ Eq. (\ref{0a13}), and a decaying wave with the
structure $u_{2}^{>}$, Eq. (\ref{0a29}). We describe the scattering process
for $\varepsilon + \varepsilon_{F} > 0$ and match these functions at the interface :%

\begin{equation}
u_{1U_{0}}^{R}+r_{22}\left( \varepsilon \right)
u_{1U_{0}}^{L}+Bu_{2U_{0}}^{<}=t_{21}\left( \varepsilon \right)
u_{1}^{R}+Cu_{2}^{>},  \label{0a31}
\end{equation}%

Eq. (\ref{0a31}) presents a system of $4$ linear equations. They can be
simplified using the fact that $U_{0}\gg \varepsilon, \varepsilon_{F}$ and the
amplitudes $t_{21}\left( \varepsilon \right) $ and $r_{22}\left( \varepsilon
\right) $ can be found denoting :%

\begin{equation}
\frac{K_{0}}{K_{0}U_{0}}\simeq \sqrt{\frac{\left\vert \varepsilon
+ \varepsilon_{F}\right\vert }{U_{0}}}=L\left( \varepsilon \right) \ll 1  \label{0a32}
\end{equation} %

we obtain using this approximation :%

\begin{equation}
t_{21}\left( \varepsilon \right) =2\sqrt{L\left( \varepsilon \right) \cos
\alpha \left( 1+\sin ^{2}\alpha \right) }\exp \left( -i\Phi \left( \alpha
\right) \right) ,  \label{0a33}
\end{equation}%

where the angle $\Phi \left( \alpha \right) $ varies in the interval $0<\Phi \left( \alpha \right) <\pi /2 $ :%

\begin{equation}
\Phi \left( \alpha \right) =\arcsin \left( \sin ^{2}\alpha \right) .
\label{0a34}
\end{equation}%

We obtain for the reflection coefficient :%

\begin{equation}
r_{22}\left( \varepsilon \right) =1-2L\left( \varepsilon \right) \sqrt{%
\left( 1+\sin ^{2}\alpha \right) }\exp \left( -i\Phi \left( \alpha \right)
\right)  \label{0a36}
\end{equation}%

with the unitarian condition :%

\begin{equation}
\left\vert t_{21}\left( \varepsilon \right) \right\vert ^{2}+\left\vert
r_{22}\left( \varepsilon \right) \right\vert ^{2}=1  \label{0a37}
\end{equation}%

that immediately follows from Eqs. (\ref{0a33}, \ref{0a36}).

\paragraph{Region $\protect\varepsilon + \varepsilon_{F} < 0.$}

In the region $\varepsilon + \varepsilon_{F} < 0$ matching the wave functions at $x=0$
results in the equation :%

\begin{equation}
u_{1U_{0}}^{R}+r_{22}\left( \varepsilon \right) u_{1U_{0}}^{L}+B\left(
\varepsilon \right) u_{2U_{0}}^{<}=t_{21}\left( \varepsilon \right)
u_{2}^{R}+C\left( \varepsilon \right) u_{1}^{>},  \label{0a38}
\end{equation}

Using the same approximation as in (\ref{0a32}) and proceeding in the same way as
for $\varepsilon + \varepsilon_{F} $ one comes to the following results, that again fulfil the unitarian condition in Eq. (\ref{0a37}) :%

\begin{equation}
t_{21}\left( \varepsilon \right) =-2\sqrt{L\left( \varepsilon \right) \cos
\alpha ^{\prime }}\sin \alpha ^{\prime }\exp \left( i\Phi \left( \alpha
^{\prime }\right) \right)  \label{0a39}
\end{equation}%

with the reflection amplitude $r_{22}\left( \varepsilon \right) $ :%

\begin{equation}
r_{22}\left( \varepsilon \right) =1-2iL\left( \varepsilon \right) \cos
\alpha ^{\prime }\exp \left( i\Phi \left( \alpha ^{\prime }\right) \right) ,
\label{0a40}
\end{equation}%

and :%

\begin{equation}
\Phi \left( \alpha ^{\prime }\right) =-\arcsin \left( \sin ^{2}\alpha
^{\prime }\right)  \label{0a41}
\end{equation}%

It is well known~\cite{LL} that the transmission amplitude $%
t_{12}\left( \varepsilon \right) $ is related to the amplitude $t_{21}\left(
\varepsilon \right) $ as :%

\begin{equation}
t_{12}\left( \varepsilon \right) =t_{21}\left( \varepsilon \right)
e^{i\delta \left( \varepsilon \right) },  \label{0a42}
\end{equation}%

where no knowledge of the explicit form of the phase $\delta \left( \varepsilon \right) $ is necessary for the calculation of the
conductances.

\subsection{Differential conductance of BLG.}

Using Eqs. (\ref{0a7}-\ref{0a10}), the calculations of the conductances $%
G_{ns}$ and $G_{nn}$ are straightforward. From the calculated $t_{21}\left( \varepsilon \right) $ and $r_{22}\left( \varepsilon
\right) $ one can easily obtain $t_{21}\left( -\varepsilon
\right) $ and $r_{22}\left( -\varepsilon \right) $. We concentrate now on the case $\varepsilon > 0$, $\varepsilon_{F} > 0$ where we can distinguish two regions in the parameter space, where $\varepsilon < \varepsilon_{F}$ and the reflections are of the retro type, and $\varepsilon > \varepsilon_{F}$ where the reflections are of the specular type.

\subsubsection{ Explicit expressions for the conductance $G_{ns}\left(
\protect\varepsilon \right) $ at $\protect\varepsilon < \varepsilon_{F}$ (retro
reflection).}

Eqs. (\ref{0a7}-\ref{0a10}) can be rewritten using the integration over $k_{y}$ instead of the trace over
the transversal channels. However, it is even more convenient to integrate over the
incident angle $\alpha $. The angle $\alpha $ corresponds to the energy $%
\varepsilon$, while another angle $\alpha ^{\prime }$ corresponds to the
energy $-\varepsilon $. These angles are related to each other by the
condition that $k_{y}$ is the same in both the cases :%

\begin{equation}
\frac{\sin \alpha ^{\prime }}{\sin \alpha }=-\frac{K_{0}\left( \varepsilon
\right) }{K_{0}\left( -\varepsilon \right) }=-\sqrt{\frac{\varepsilon_{F}+\varepsilon
}{\varepsilon_{F}-\varepsilon }}  \label{0a43}
\end{equation}%

resulting in the condition that Andreev reflections are only possible for angles $\left\vert \alpha \right\vert <\alpha _{c},$ where :%

\begin{equation}
\alpha _{c}=\arcsin \sqrt{\frac{\varepsilon_{F}-\varepsilon }{\varepsilon_{F}+\varepsilon }}
\label{0a44}
\end{equation}%

With :%

\begin{equation}
\Phi \left( \alpha \right) =\arcsin \left( \sin ^{2}\alpha \right) ,
\label{0a45}
\end{equation}%

and :%

\begin{equation}
\Phi \left( \alpha ^{\prime }\right) =\arcsin \left( \sin ^{2}\alpha
^{\prime }\right) =\arcsin \left[ \left( \frac{\varepsilon_{F}+\varepsilon }{%
\varepsilon_{F}-\varepsilon }\right) \sin ^{2}\alpha \right]  \label{0a45a}
\end{equation}%

the conductance $G_{ns}$ can be reduced to the form :%

\begin{equation}
G_{ns}=2G_{0}K_{0}\left( \varepsilon \right) \int_{0}^{\alpha _{c}}\frac{%
Y_{\varepsilon }\left( \alpha ,\alpha ^{\prime }\right) \cos \alpha }{%
2\left\vert X_{\varepsilon }\left( \alpha ,\alpha ^{\prime }\right)
\right\vert ^{2}}d\alpha  \label{0a46}
\end{equation}

where :%

\begin{eqnarray}
&&Y_{\varepsilon }\left( \alpha ,\alpha ^{\prime }\right) =\left\vert
t_{21}\left( \varepsilon \right) \right\vert ^{2}\left\vert t_{21}\left(
-\varepsilon \right) \right\vert ^{2}=  \label{0a47} \\
&&=16L\left( \varepsilon \right) L\left( -\varepsilon \right) \cos \alpha
\cos \alpha ^{\prime }\left( 1+\sin ^{2}\alpha \right) \left( 1+\sin
^{2}\alpha ^{\prime }\right)  \notag
\end{eqnarray}%

and:%

\begin{equation}
X_{\varepsilon }\left( \alpha ,\alpha ^{\prime }\right) =\frac{1}{2}\left[
1-e^{-2i\beta }r_{22}^{\ast }\left( -\varepsilon \right) r_{22}\left(
\varepsilon \right) \right]  \label{0a48}
\end{equation}%

Here we can write $\left\vert X_{\varepsilon }\right\vert ^{2}$ entering Eq. (\ref%
{0a47}) only up to quadratic terms in $L\left( \varepsilon \right) $ :%

\begin{eqnarray}
&&\left\vert X_{\varepsilon }\left( \alpha ,\alpha ^{\prime }\right)
\right\vert ^{2}=\sin ^{2}\beta - \label{0a50} \\
&&-2\sin \beta \Big[L\left( \varepsilon
\right) \sqrt{1+\sin ^{2}\alpha }\sin \left( \beta +\Phi \left( \alpha
\right) \right) + \notag \\
&&+L\left( -\varepsilon \right) \sqrt{1+\sin ^{2}\alpha ^{\prime }}\sin
\left( \beta -\Phi \left( \alpha ^{\prime }\right) \right) \Big]+  \notag \\
&&+L^{2}\left( \varepsilon \right) \left( 1+\sin ^{2}\alpha \right)
+L^{2}\left( -\varepsilon \right) \left( 1+\sin ^{2}\alpha ^{\prime }\right)+
\notag \\
&&+2L\left( \varepsilon \right) L\left( -\varepsilon \right) \sqrt{\left(
1+\sin ^{2}\alpha \right) \left( 1+\sin ^{2}\alpha ^{\prime }\right) }\times
\notag \\
&&\times \Big[\cos \left( 2\beta -\Phi \left( \alpha ^{\prime }\right) +\Phi
\left( \alpha \right) \right) -2\sin \Phi \left( \alpha \right) \sin \Phi
\left( \alpha ^{\prime }\right) \Big]  \notag
\end{eqnarray}%

\subsubsection{Explicit expressions for the conductance $G_{ns}\left(
\protect\varepsilon \right) $ at $\protect\varepsilon > \varepsilon_{F}$ (specular
reflection).}

Here the angles $\alpha $ and $\alpha ^{\prime }$ are related to
each other as :%

\begin{equation}
\frac{\sin \alpha ^{\prime }}{\sin \alpha }=\frac{K_{0}\left( \varepsilon
\right) }{K_{0}\left( -\varepsilon \right) }=\sqrt{\frac{\varepsilon + \varepsilon_{F}}{%
\varepsilon - \varepsilon_{F}}}  \label{0a51}
\end{equation}%

and the critical angle $\alpha _{c}$ equals :%

\begin{equation}
\alpha _{c}=\arcsin \sqrt{\frac{\varepsilon - \varepsilon_{F}}{\varepsilon + \varepsilon_{F}}}
\label{0a52}
\end{equation}%

For $\Phi \left( \alpha \right) $ we use Eq. (\ref{0a45})
and obtain for $\Phi \left( \alpha ^{\prime }\right) $ :%

\begin{equation}
\Phi \left( \alpha ^{\prime }\right) =\arcsin \left[ \left( \frac{%
\varepsilon + \varepsilon_{F}}{\varepsilon - \varepsilon_{F}}\right) \sin ^{2}\alpha \right]
\label{0a53}
\end{equation}%
\newline

\paragraph{Conductance for $\protect\varepsilon > \varepsilon_{F}.$}

In this case the conductance $G_{ns}\left( \varepsilon \right) $ is
determined by Eq. (\ref{0a46}). Therefore we obtain :%

\begin{eqnarray}
&&Y_{\varepsilon }\left( \alpha ,\alpha ^{\prime }\right) =\left\vert
t_{21}\left( \varepsilon \right) \right\vert ^{2}\left\vert t_{21}\left(
-\varepsilon \right) \right\vert ^{2} = \label{0a54} \\
&&=16L\left( \varepsilon \right) L\left( -\varepsilon \right) \cos \alpha
\left( 1+\sin ^{2}\alpha \right) \cos \alpha ^{\prime }\sin ^{2}\alpha
^{\prime }  \notag
\end{eqnarray}%

and as $X_{\varepsilon }\left( \alpha ,\alpha ^{\prime }\right) $ is
determined by Eq. (\ref{0a48}) we get :%

\begin{eqnarray}
&&\left\vert X_{\varepsilon }\left( \alpha ,\alpha ^{\prime }\right)
\right\vert ^{2}=\sin ^{2}\beta - \label{0a56} \\
&&-2\sin \beta \Big[L\left( \varepsilon
\right) \sqrt{1+\sin ^{2}\alpha }\sin \left( \beta +\Phi \left( \alpha
\right) \right) - \notag \\
&&-L\left( -\varepsilon \right) \cos \alpha ^{\prime }\cos \left( \beta
+\Phi \left( \alpha ^{\prime }\right) \right) \Big]+L^{2}\left( \varepsilon \right) \left( 1+\sin ^{2}\alpha \right)+ \notag \\
&&+L^{2}\left( -\varepsilon \right) \cos ^{2}\alpha ^{\prime }+2L\left(
\varepsilon \right) L\left( -\varepsilon \right) \cos \alpha ^{\prime }\sqrt{1+\sin ^{2}\alpha }\times
\notag \\
&&\times \Big[\sin \left( 2\beta +\Phi \left( \alpha \right)
+\Phi \left( \alpha ^{\prime }\right) \right) -2\cos \phi \left( \alpha
\right) \sin \Phi \left( \alpha ^{\prime }\right) \Big]  \notag
\end{eqnarray}%

\subsubsection{Conductance $G_{nn}$ of the interface between two normal
metals.}

The transmission amplitude $t_{12}\left( \varepsilon \right) $ determines
the conductance $G_{nn}$ between two normal metals. At a fixed $\varepsilon_{F}$ the
conductance $G_{nn}\left( \varepsilon \right) $ is given by the following
formula :%

\begin{equation}
G_{nn}\left( \varepsilon \right) =G_{0}K_{0}\left( \varepsilon \right)
\int_{-\pi /2}^{\pi /2}\left\vert t_{12}\left( \varepsilon \right)
\right\vert ^{2}\cos \alpha d\alpha   \label{0a57}
\end{equation}%

which with Eq. (\ref{0a33}) results in :%

\begin{eqnarray}
G_{nn}\left( \varepsilon \right) &=&G_{0}K_{0}\left( \varepsilon \right)
\int_{-\pi /2}^{\pi /2}4L\left( \varepsilon \right) \cos ^{2}\alpha \left(
1+\sin ^{2}\alpha \right) d\alpha  \notag \\
&=&\frac{5\pi }{2}G_{0}K_{0}\left( \varepsilon \right) L\left( \varepsilon
\right)  \label{0a58}
\end{eqnarray}%

The conductance between of the normal metals can conveniently normalize the
conductance $G_{ns}$ and allows to compare with the experimental quantities :%

\begin{equation}
\bar{G}_{ns}\left( \varepsilon \right) =\frac{G_{ns}\left( \varepsilon
\right) }{\left\langle G_{nn}\left( \varepsilon ,\varepsilon_{F},T\right)
\right\rangle _{F}}  \label{0a59}
\end{equation}%

It is important to emphasize that, although the conductance $G_{ns}\left(
\varepsilon \right) $ is symmetric with respect to the inversion $%
\varepsilon \rightarrow -\varepsilon $, the normal conductance $\bar{G}_{ns}$
is not. Therefore, the reduced conductance $\bar{G}_{ns}\left( \varepsilon
\right) $ is not symmetric either. \newline

\end{document}